\documentclass[
  journal=pasa,
  manuscript=research-paper, %% or "review"
  year=2020,
  volume=37,
]{cup-journal}

\usepackage{microtype,siunitx,booktabs}
\usepackage{hyperref}  %hyperref
\usepackage{amsmath}
\usepackage{amssymb}
\usepackage{float}
\usepackage{makecell}
\usepackage{booktabs}
\usepackage{rotating}
\usepackage{xcolor}
\usepackage{graphicx}
\usepackage{afterpage}
\usepackage{subcaption}

\title{A catalogue of complex radio sources in the Rapid ASKAP Continuum Survey created using a Self-Organising Map}

\author{A. Alam}
\affiliation{E.A. Milne Centre for Astrophysics, University of Hull, Cottingham Road, Kingston-upon-Hull, HU6 7RX, UK}
\email[A. Alam]{a.alam-2019@hull.ac.uk}

\author{K.~A. Pimbblet}
\affiliation{E.A. Milne Centre for Astrophysics, University of Hull, Cottingham Road, Kingston-upon-Hull, HU6 7RX, UK}
\alsoaffiliation[2]{Centre of Excellence for Data Science, AI, and Modelling (DAIM), University of Hull, Cottingham Road, Kingston-upon-Hull, HU6 7RX, UK}

\author{Y.~A. Gordon}
\affiliation{Department of Physics, University of Wisconsin-Madison, 1150 University Ave, Madison, WI 53706, USA}

\doi{}

\received {dd Mmm YYYY}
\revised  {dd Mmm YYYY}
\accepted {dd Mmm YYYY}
\published{dd Mmm YYYY}


\begin{document}

\begin{abstract}
Next generations of radio surveys are expected to identify tens of millions of new sources, and identifying and classifying their morphologies will require novel and more efficient methods. Self-Organising Maps (SOMs), a type of unsupervised machine learning, can be used to address this problem. We map 251,259 multi-Gaussian sources from Rapid ASKAP Continuum Survey (RACS) onto a SOM with discrete neurons. Similarity metrics, such as Euclidean distances, can be used to identify the best-matching neuron or unit (BMU) for each input image. We establish a reliability threshold by visually inspecting a subset of input images and their corresponding BMU. We label the individual neurons based on observed morphologies and these labels are included in our value-added catalogue of RACS sources. Sources for which the Euclidean distance to their BMU is $\lesssim$ 5 (accounting for approximately 79$\%$ of sources) have an estimated $>90\%$ reliability for their SOM-derived morphological labels. This reliability falls to less than 70$\%$ at Euclidean distances $\gtrsim$ 7. Beyond this threshold it is unlikely that the morphological label will accurately describe a given source. Our catalogue of complex radio sources from RACS with their SOM-derived morphological labels from this work will be made publicly available.
\end{abstract}

%%%%%%%%%%%%%%%%%%%%%%%%%%%%%%%%%%%%%%%%
\section{Introduction}
\label{Intro}

Astronomical radio emission is dominated by synchrotron emission resulting from charged particles moving at relativistic velocities through magnetic fields. Such synchrotron radiation in galaxies generally arise due to the supermassive black hole at its centre or remnants from supernovae. Extragalactic radio continuum surveys therefore generally detect two groups of galaxies: star-forming galaxies (\citealp{Condon}) and active galactic nucleus (AGN; \citealp{Kormendy}). While AGN emit emission across the entire electromagnetic spectrum (\citealp{Padovani2017}), a fraction of them (around 15-20$\%$) are considered radio-loud and produce strong radio emission as a result of synchrotron emission from the AGN's relativistic jets (\citealp{Sadler}, \citealp{Kellermann1989}, \citealp{Urry1995}). Early radio astronomers faced difficulties in detecting radio galaxies at optical wavelengths since they are very faint at such wavelengths and this meant that there was little overlap between optical and radio surveys (\citealp{SavageWall}, \citealp{Windhorst1984}, \citealp{Windhorst1985}).

The advent of new technology led to a transformational period in radio astronomy between 1990 and 2004 where surveys such as Westerbork Northern Sky Survey (WENSS; \citealp{WENSS}), NRAO VLA Sky Survey (NVSS; \citealp{NVSS}), Faint Images of the Radio Sky at Twenty-Centimeters (FIRST; \citealp{FIRST}), and Sydney University Molonglo Sky Survey (SUMSS; \citealp{SUMSS}) enabled an increase of the number of known radio sources to around 2.5 million sources. This is a hundredfold increase from earlier surveys such as 3C (\citealp{3C}) and Parkes Catalogue of Radio Sources (\citealp{Parkes}) among others (see summary in \citealp{Norris2017}). However, even with improved sensitivities at radio-wavelengths, radio catalogues from these surveys are still $\sim$1$\%$ in size relative to those at optical (\citealp{Norris2017}).

Next-generation radio continuum surveys are expected to go further and observe tens of millions of new objects (\citealp{Norris2017}). This includes 20$\%$ of galaxies which were detected by infrared surveys such as Wide-field Infrared Survey Explorer (WISE; \citealp{Wright}) and multi-spectral imaging and spectroscopic redshift surveys such as Sloan Digital Sky Survey (SDSS; \citealp{York2000}, \citealp{SDSS14}). Telescopes such as the Square Kilometre Array (SKA) and its precursor and pathfinder instruments such as Australian Square Kilometre Array Pathfinder (ASKAP; \citealp{Johnston2008}, \citealp{Hotan}), Low Frequency Array (LOFAR; \citealp{vanHaarlem}), and Murchison Widefield Array (MWA; \citealp{Tingay2013}, \citealp{Wayth2018}), in addition to the Karl G. Jansky Very Large Array (JVLA; \citealp{VLA}) are able to conduct deep continuum surveys that can detect millions of radio sources. Moreover, they would be able to carry out these surveys on much shorter timescales than earlier surveys such as the LOFAR Two-metre Sky Survey (LoTSS; \citealp{LoTSS}), the Evolutionary Map of the Universe Pilot Survey (EMU; \citealp{EMU2}, \citealp{Emupilot}), and the Very Large Array Sky Survey (VLASS; \citealp{VLASS}). 

Radio sources with a single component, where component refers to an output (generally a 2D Gaussian) from a source finding algorithm, are termed simple sources and they tend to make up around the majority of radio sources (\citealp{Norris2017}). They can be easily resolved and cross-matched with catalogues in different wavelengths such as optical and infrared using techniques such as the Likelihood Ratio method (\citealp{Sutherland}). Complex sources, which are expected to make up a significantly smaller fraction of radio sources, are those which have multiple radio components and cannot be as easily identified as simple sources (\citealp{Williams2019}, \citealp{Gurkan2022}, \citealp{Gordon2023}). For example, two unresolved radio components that are in close proximity to each other might be radio emission from separate galaxies or they might be the lobes of a radio source (\citealp{Gordon2023}). 

Given the large number of sources that are expected to be observed by next-generation surveys, classifying the morphology of the detected radio sources will be a challenging undertaking that will require novel methods of cross-identification. In this paper, we explore the use of self-organising maps (SOM; \citealp{Kohonen1990}, \citealp{Kohonen}), an unsupervised machine learning algorithm, in order to address the problem of finding complex radio sources and classifying their morphologies in the large dataset provided by SKA pathfinders, such as the Rapid ASKAP Continuum Survey (RACS; \citealp{McConnell2020}). \citet{Galvin2020} used SOMs to identify related radio components and the corresponding infrared host galaxy. The SOM was used on 894,415 images from FIRST and infrared data from WISE (\citealp{Wright}) centred at positions described by the FIRST catalogue. Using a SOM, they were able to identify potentially resolved radio components which correspond to a single infrared host. Moreover, their approach was able to detect radio objects with interesting and rare morphologies such as `X'-shaped galaxies, and introduce a statistic that will enable the search of bent and disturbed radio morphologies. By using their method, they were able to identify 17 giant radio galaxies between 700 - 1100 kpc. 

The layout of the rest of this paper is as follows: in Section \ref{Data} we provide a brief description of the RACS data products used in this paper, and in Section \ref{method} we outline the preprocessing steps for the cutout images from RACS as well as the SOM training process. In Section \ref{inspection&mapping} we analyse the results from the SOM training, and describe the inspection and mapping done to create a catalogue of complex sources, and we summarise our conclusions in section \ref{Conclusions}. This is followed by an \ref{appendix1} containing additional figures exploring the properties of individual neurons in the trained SOM grid.

%%%%%%%%%%%%%%%%%%%%%%%%%%%%%%%%%%%%%%%%
\section{Data}
\label{Data}

The survey used in this paper is from the first epoch of RACS, which is the first all-sky survey conducted with the full ASKAP telescope (\citealp{McConnell2020}, \citealp{Hale2021}). ASKAP is an array of 36 antennas, each with a 12-meter diameter. Each antenna is equipped with a phased array feed (PAF) and is capable of dual polarisation. At $800\,$MHz, each ASKAP pointing has a field of view of $\approx 31\,\text{deg}^{2}$ (\citealp{Hotan}). RACS is the deepest radio survey covering the entire southern sky to date. It is able to connect low-frequency surveys such as TIFR GMRT Sky Survey (TGSS; \citealp{TGSS}) and Galactic and Extragalactic Allsky Murchison Widefield Array survey (GLEAM; \citealp{GLEAM}) to surveys such as NVSS (\citealp{NVSS}) at 1.4 GHz and VLASS at 3 GHz (\citealp{VLASS}). 

We specifically use the data products from the first public data release from RACS, RACS-Low, which is made up of 903 tiles south of declinations of +41° and covered a total survey area of $34240\,\text{deg}^{2}$. It is centred at 887.5 MHz, with 15-minute integrations and 288 MHz of bandwidth with 1 MHz wide channels, and they achieved a nominal sensitivity between 0.25-0.3mJy/beam (\citealp{McConnell2020}, \citealp{Hale2021}). The first Stokes I catalogue from \citet{Hale2021} is derived from 799 tiles that have been convolved to a common resolution of 25 arcseconds. It covered most of the sky in the declination region $\delta = -80$° to $+30$°, excluding the region $|$ b $| < 5$° in the Galactic plane. The catalogue uses Python Blob Detection and Source Finder (\texttt{PyBDSF}; \citealp{PyBDSF}) to detect regions of radio emission which are fitted with 2D Gaussian components. As such, the catalogue contains both single source components as well as sources with multiple components which are defined as single sources or islands of pixels fitted with multiple Gaussians. The catalogue contains 2,123,638 sources, of which 1,872,361 (88.17\%) are simple sources with a single Gaussian component and 251,277 sources (11.83\%) are complex with multiple components. Since the objective of our work is to identify complex sources, the focus will be on this latter group of sources with multiple components.

\section{Methodology}
\label{method}

\subsection{Self-Organising Map (SOM)}
\label{soms}

Given that the next-generation surveys are expected to become more data intensive and detect vast number of radio sources, machine learning methods can help us identity and classify these detected sources in a more efficient and labour-saving manner. They can be divided into two approaches: supervised and unsupervised machine learning. Supervised machine learning broadly describe algorithms that learn to represent an unknown and possibly complex function by training a mapping function between input data and their assigned training labels. For example, \citet{Aniyan} used convolution neural network (CNN), a type of supervised machine learning method, to classify radio sources from FIRST into Fanaroff-Riley (\citealp{Fanaroff&Riley}) and bent-tailed morphology classes, and had a success rate of approximately 95$\%$ depending on the morphology presented, with bent-tailed radio galaxies being the most identifiable. \citet{Lukic2018} used CNNs trained on radio sources from Radio Galaxy Zoo (RGZ; \citealp{Banfield}) Data Release 1 to classify sources into compact and different classes of extended sources and achieved a 94.8$\%$ accuracy rate.

Conversely, unsupervised machine learning methods are algorithms that do not need labelled data or any prior knowledge about the dataset. Instead these algorithms focus on identifying any existing structures within a dataset. Examples of unsupervised machine learning methods include k-means clustering (\citealp{kmeans}), Gaussian mixture models (\citealp{GMM}), principal component analysis (\citealp{PCA}), and self-organising maps (SOMs; \citealp{Kohonen1990}, \citealp{Baron}). SOMs (also known as Kohonen maps) are a type of neural network which output a low-dimensional, usually two-dimensional, representation of the input dataset. SOMs use a competitive learning process in order to map the input dataset onto a grid of discrete neurons. The neurons are each assigned a unique position, $i$, onto a regular lattice and initialised with prototype weights, $w$, which are usually zeros or small random values. A single iteration of training involved randomly selecting an input data sample, $d$, from the reference training dataset, $D$, and comparing it to the current state of each of the neurons. The neuron with the best similarity score is referred to as the Best-Matching Unit (BMU). The position of the BMU, $j$, is then used to update the weights of the other neurons in the grid. 

Euclidean distance is one of the similarity metrics that the SOM algorithm can use to quantify the similarity between the input data and the neurons in the SOM grid. This can be done by calculating the straight-line distances between them. However, this can result in problems with some data types, including astronomical images, which do not maintain invariance between certain types of transformations such as flipping and rotation. As a result, \citet{Polsterer} developed the software Parallelized rotation and flipping INvariant Kohonen-maps (\texttt{PINK})\footnote{Parallelized rotation and flipping INvariant Kohonen maps (\texttt{PINK}): \url{https://github.com/HITS-AIN/PINK}} which builds upon the SOM algorithm and introduces a minimisation procedure to best align a source of random orientation onto the neurons. This ensures that sources which are similar are grouped despite any differences in rotation. A SOM training using \texttt{PINK} starts with the weights of all the neurons being initialised with randomly generated numbers or zeros. \texttt{PINK} then rotates and flips all input images a set number of times (specified by the Rotations parameter which will be discussed in more detail in \ref{training}). The similarity between the input images, including the rotated and flipped copies, and all the neurons are calculated so that the BMU, i.e. the neuron with the shortest Euclidean distance to a given input image and is therefore the best representation of the input, can be identified (see Section 2 of \citealp{Polsterer} for more details on the method used in the \texttt{PINK} framework, and \citealp{Polsterer2015} for details on an earlier version of the framework). The weighting function implemented in \texttt{PINK} is described by Equation \ref{eq:som} (\citealp{Polsterer}): 
\begin{equation} 
w'_i = w_i + \alpha(t) \cdot G_{ij} \cdot (\phi(d) - w_i)
\label{eq:som}
\end{equation}
where:
\begin{itemize}
    \item \(w_i\) is the initial weight vector of neuron \(i\).
    \item \(w'_i\) is the updated weight vector of neuron \(i\).
    \item \(\alpha\) is the learning rate and this parameter controls how much the weights are updated as training progresses.
    \item \(G_{ij}\) is the neighbourhood function which controls the extent to which the BMU neuron \(j\) influences the weight update of neuron \(i\). \texttt{PINK} currently supports three possible neighbourhood functions of which the Gaussian distribution function is the most common and is used for this training. The width of the Gaussian neighbourhood function establishes the BMU's region of influence such that the weights of the neurons which are closer to the BMU are updated more than neurons which are further away.  
    \item \(d\) is the current input data, and the term \((\phi(d) - w_i)\) aligns \(d\) onto \(w_i\).
\end{itemize}

\texttt{PINK} uses a modified Euclidean distance metric (\citealp{Polsterer}, \citealp{Galvin2019}) to measure similarity:
\begin{equation}
\Delta(A, B) = \underset{\forall \phi \in \Phi}{\text{minimize}(\phi)} \sqrt{\sum_{c=0}^{C} \sum_{x=0}^{X} \sum_{y=0}^{Y} \left( A_{c,x,y} - \phi(B_{c,x,y}) \right)^2}
\label{eq:modified_euclidean}
\end{equation}
where $A$ and $B$ are a given neuron and input image and $c$ is their corresponding channel, $x$ and $y$ are the coordinates of the pixels, $\phi$ corresponds to an affine image transformation which has been drawn from a set of image transformations $\Phi$, i.e. the set of all rotated and flipped copies of the input, and is optimised to best align the input image with the features of the neuron by finding the $\phi$ with the shortest Euclidean distance from all the possible rotations and flips. \texttt{PINK} can also impose either a circular or quadratic region over which the Euclidean distances are calculated. A quadratic region can cause variations in these calculations due to the impact of other sources, especially bright sources, potentially moving into the masked region as the images are rotated (\citealp{Vantyghem}). We use a circular mask in order to minimise the effects of sources near the edges of the images. Once the Euclidean distances have been calculated and a BMU has been determined, the neuron positions on the SOM grid are evaluated against the neighbourhood function and their weights are modified accordingly so that they can be a better representation of the input. The previous steps are then iterated over all input images in the training dataset an $X$ amount of times, keeping in consideration that $X$ must be large enough to allow the SOM to converge. After enough training iterations have taken place such that stable SOM can be produced, all input images in the dataset are mapped to the derived neurons in order to determine the distances to the neurons and find the best match regions. 

\begin{figure*}[ht!]
    \centering
    \includegraphics[width=\textwidth]{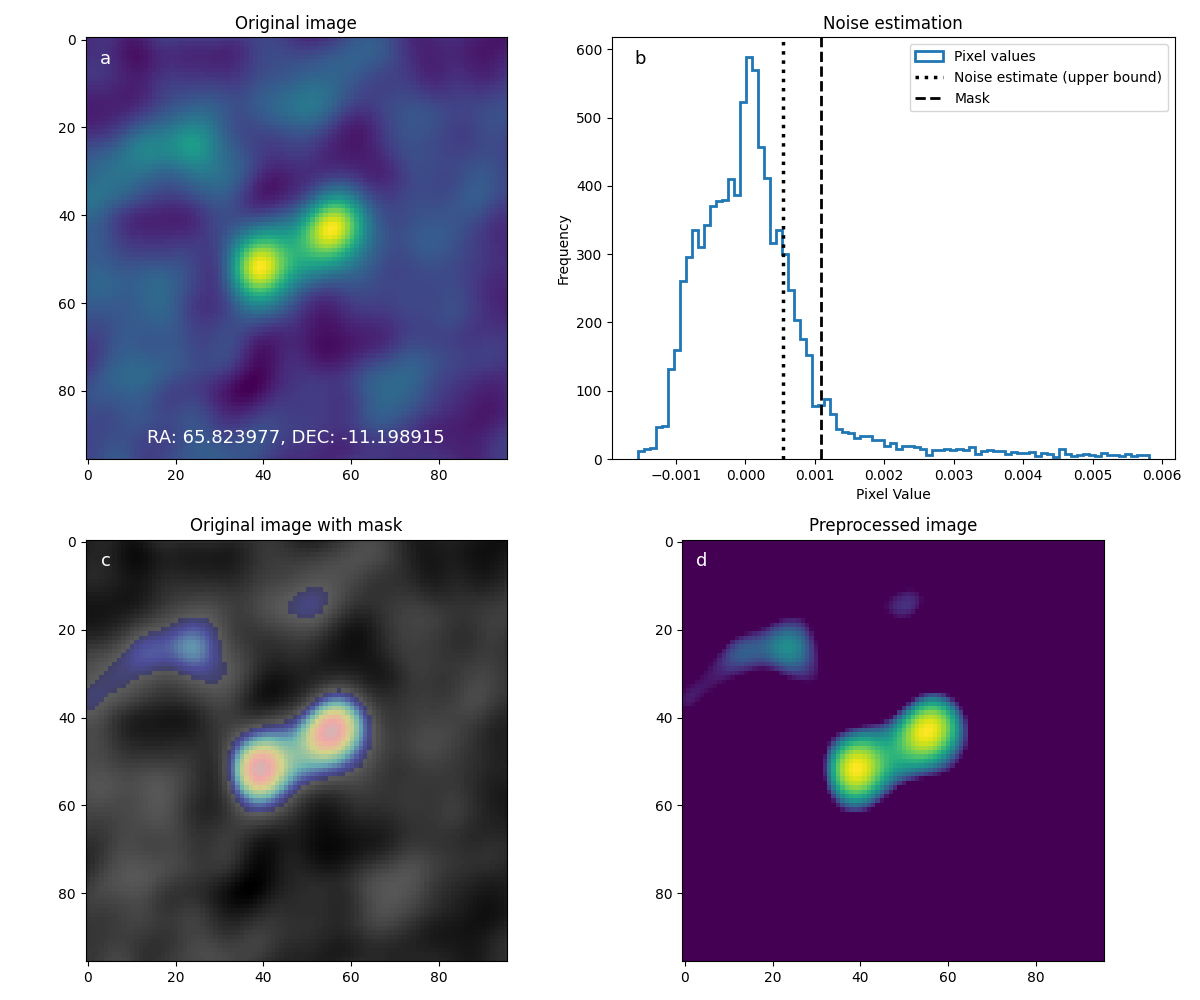}
    \caption{The preprocessing stages for each of the RACS image cutouts for a randomly chosen image cutout. On the first panel (a) we have the original image with its RA and DEC coordinates. On the second panel (b) we show the distribution of the pixel values along with the upper bound of the noise estimate and the mask applied (given by noise estimate multiplied by a minimum signal-to-noise value of 2). On the third panel (c) we show the original image with the mask overlaid on top. Once a mask is applied, we log scale the remained pixels and normalise them from 0 to 1 which gives us the final preprocessed image in the fourth panel (d).} 
    \label{fig:preprocessing_stages}
\end{figure*} 

The key hyperparameters to consider during the training process are: the number of neurons in the SOM grid, width of the neighbourhood function, learning rate, and the number of rotations and iterations. The number of neurons specifies the size of the SOM, and should be large enough to represent the dominant structures in the training data but not so large that it becomes time-consuming to compute. The width of the neighbourhood function used cannot be too wide as that would impact all neurons in the grid, however it also cannot be too narrow as that could potentially lead to individual neurons being decoupled and only creating smaller clusters that do not accurately capture the similarities between neighbouring neurons. The learning rate, if too small, will result in long computational time, but if it is too large then the changes in the weight updates would be too abrupt (\citealp{Mostert2021}).

\subsection{Preprocessing RACS image cutouts}
\label{preprocessingimages}

Prior to training a SOM, the data is preprocessed to ensure that the training set emphasises the morphological structures and features present and interference from noise is minimised. The first step is to filter the RACS-Low catalogue to select complex sources with multiple components which are classed as `multi-Gaussian' (given by `M' in column `S Code' in the catalogue) by \texttt{PyBDSF} (\citealp{PyBDSF}, \citealp{Hale2021}). This results in 251,277 individual radio sources. Next, we take 96 x 96 pixel cutouts from the RACS image tiles centred around each coordinate which corresponds to a \ang{;4;} field-of-view. The vast majority of radio sources are expected to have sub-arcmin sizes, with only the largest sources having angular extents greater than \ang{;4;} (\citealp{Lara2001}, \citealp{Proctor2016}). A cutout of angular size \ang{;4;} will be large enough to capture most radio sources, including possible extended structures, while still being small enough to avoid any interference from unrelated nearby sources in the tile. There may exist more than one coordinate for a given object due to decomposition, hence the same object can be in more than one cutout. As such, if there are multiple cutouts for a given `target' RA and Dec coordinates, we choose the cutout that is closest to this target. This is done by calculating the angular distance between the pixel coordinates of the target and the reference CRPIX1 and CRPIX2 pixel coordinates of the cutouts. The most central cutout, i.e. the cutout with the smallest angular distance to the reference pixel coordinates, is selected.

\begin{figure}[!h]
    \centering    
    \includegraphics[width=\columnwidth]{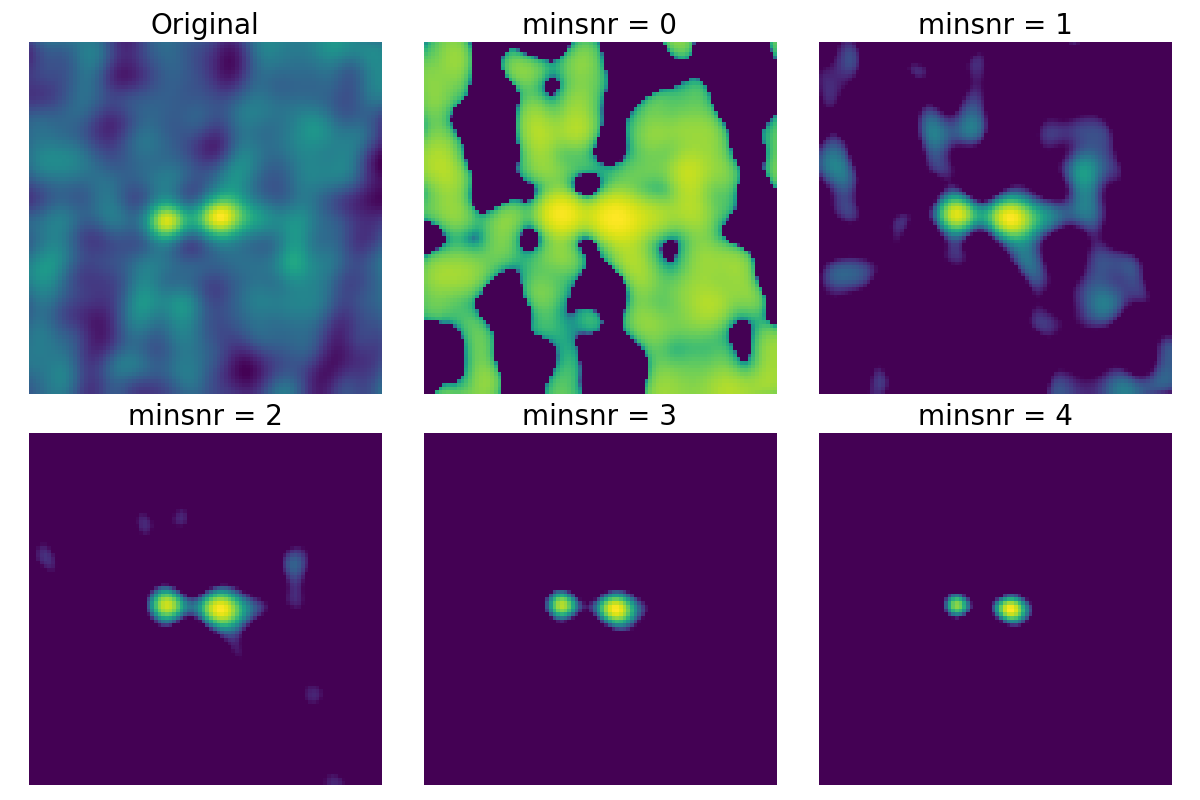}  
    \caption{The original cutout and the preprocessed images of the cutouts done with different values of the minimum signal-to-noise ratio (0, 1, 2, 3 and 4) for the mask. The value was set to 2 since it is enough to mask the majority of the noise without losing much information as we can see in the images.}
    \label{fig:minsnr_test}
\end{figure}

For our preprocessing we use the python package \texttt{PYINK}\footnote{\texttt{PYINK}: \url{https://github.com/tjgalvin/pyink} (commit \texttt{176177b})}, which has a set of useful tools to aid in training and analysing a SOM using \texttt{PINK} and can be used to create the bespoke binary file it requires. The image cutouts for each coordinate are preprocessed with \texttt{PYINK} and Figure \ref{fig:preprocessing_stages} shows the preprocessing stages for an arbitrary cutout taken from our sample. The first stage is estimating the noise in each image cutout. We measure the outlying pixels that deviate from the median by more than three times the Median Absolute Deviation (MAD) of the pixels. We clip and remove these flagged outlier pixels and this process is repeated twice. A Gaussian was fitted to the pixel intensity distribution of the remaining unflagged pixels. The standard deviation of this fitted distribution gives the estimate of the background noise, which we can consider to be the upper bound of the noise. A mask is then created by multiplying the noise estimation and a minimum signal-to-noise ratio (the dotted and dashed lines in the second panel (b) in Figure \ref{fig:preprocessing_stages} corresponds to the noise estimation and mask respectively). Tests were run with different values of the minimum signal-to-noise (Figure \ref{fig:minsnr_test}), and based on these we determine that a value of 2 is sufficient in masking the majority of the noise in the cutout whilst still capturing the important structures present. Once a mask is applied, it filters the pixels to only retain those that satisfy the mask conditions and pixels outside this mask were filled in to have values of 0. We then apply a log scaling to the pixels and normalise them from 0 to 1 to give the final preprocessed image. Out of the 251,277 image cutouts that underwent the preprocessing method 18 cutouts failed. Visual inspections of these cutouts show that they tend to have diffuse or large scale structure which cover a larger fraction of the cutouts, and as such \texttt{PYINK} finds a higher noise level estimation than \texttt{PyBDSF}. This results in edge cases where the signal-to-noise within the cutout is not representative of the local signal-to-noise for large sources. Due to the higher noise level, the data array is empty once the mask (the product of the \texttt{PYINK} noise estimation and a minimum signal-to-noise value of 2) is applied during preprocessing. We use the remaining 251,259 sources to train the SOM. 

%%%%%%%%%%%%%%%%%%%%%%%%%%%%%%%%%%%%%%%%%%%%%

\subsection{Training}
\label{training}

Following the preprocessing, we train a 10 x 10 SOM using \texttt{PINK} with four training stages. In order to do so, we have to establish certain hyperparameters for each stage: the width of the neighbourhood function ($\sigma$), the learning rate (\(\alpha\)), and the number of rotations and iterations. As stated in Section \ref{soms}, the width of the $\sigma$ function sets how much of the neighbourhood should be updated in each iteration, the learning rate controls how much the weights are updated during each iteration, rotations gives the number of rotations and flips \texttt{PINK} performs for each input image during the training, and iterations is the number of times each individual item in the training dataset is used to update the SOM. 

\begin{table}
    \centering
    \begin{tabular}{c|c|c|c|c}        
        \textbf{Stage} & \textbf{$\sigma$} & \textbf{\(\alpha\)} & \textbf{Rotations} & \textbf{Iterations} \\%[5pt] 
        \midrule
        1 & 1.5 & 0.1 & 92 & 5 \\
        2 & 1.0 & 0.05 & 180 & 5 \\
        3 & 0.7 & 0.05 & 360 & 5 \\
        4 & 0.5 & 0.005 & 360 & 10 \\
    \end{tabular}
    \caption{The hyperparameters used in the four training stages: the width of the neighbourhood function \(G_{ij}\) given by $\sigma$, the learning rate \(\alpha\), the number of rotations and iterations.} 
    \label{table:som_parameters}
\end{table}

During the training process, we want to first establish the broad morphologies and subsequently fine-tune the SOM and identify smaller structures and details present (\citealp{Galvin2019}, \citealp{Mostert2021}). For the first training stage, the neurons are initialised with random noise. We keep the rotations and iterations to a lower number initially, and this also has the added advantage of decreasing the computational time. The number of rotations are subsequently increased at each stage, and in the final training stage we also increase the number of iterations in order to capture the finer details. \citet{Kohonen} states that a larger neighbourhood function is able to capture the broad or global structures in the dataset, and so decreasing the size of the neighbourhood function as training progresses ensures that the smaller or localised details are also represented in the SOM. We follow the same principle for the learning rate which controls the magnitude of the weight updates during each iteration. Hence, the width of the neighbourhood function ($\sigma$) and the learning rate (\(\alpha\)) are set to 1.5 and 0.1 respectively, and they are decreased in subsequent training stages. 

\begin{figure*}
    \centering
    \includegraphics[width=0.91\columnwidth]{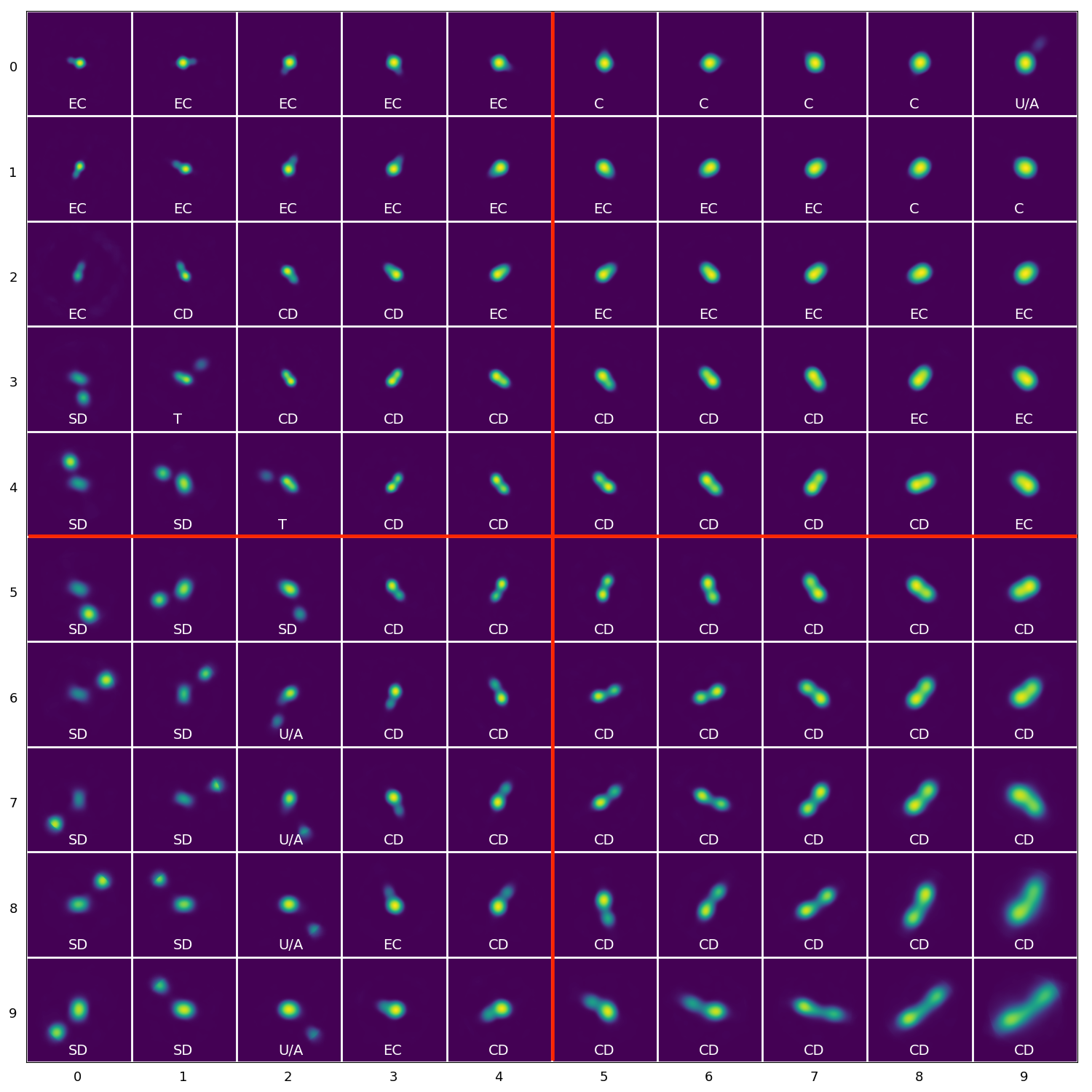}
    \caption{The trained 10x10 SOM with manual labels of their morphological labels: \textbf{C} (Compact) sources, \textbf{EC} (Extended Compact), \textbf{CD} (Connected Double) sources, \textbf{SD} (Split Double) sources, \textbf{T} (Triple) sources, \textbf{U/A} (Uncertain/Ambiguous) sources. The labels on the axis indicate the neuron coordinate in the SOM grid such that the top left neuron is (0, 0) with morphological label EC. The SOM can also divided into four quadrants: top left, top right, bottom left, and bottom right (marked in red) for additional analysis.}
    \label{fig:som_final}
\end{figure*}

The SOM grid is trained on the hyperparameters established in Table \ref{table:som_parameters} and is shown in Figure \ref{fig:som_final}. Here the neuron positions are given by (y, x) and the origin in this coordinate scheme is the left-most column of the top row, such that the right-most column of the top row has a coordinate of (0, 9). The training is done on a single GPU node with a NVIDIA A40 48GB GPU and took approximately 14 days, however we note that CUDA was disabled and this may have affected the time taken for the training.

\begin{figure}
    \centering
    \includegraphics[width=1.05\textwidth]{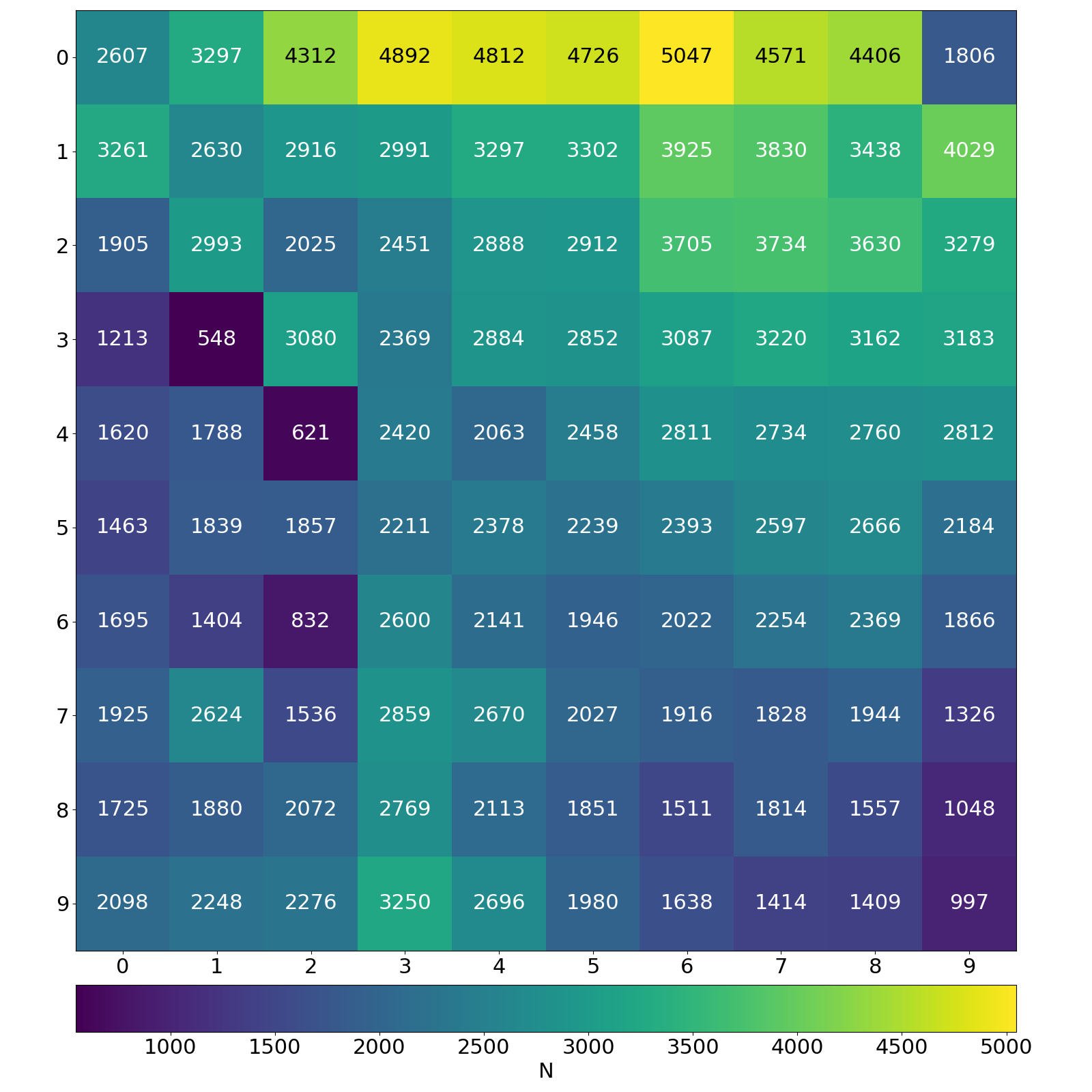}
    \caption{Density map showing the Best-Matching Unit (BMU) count across the trained SOM.}
    \label{fig:bmu_count}
\end{figure}

\section{SOM Inspection and Mapping}
\label{inspection&mapping}

Once the SOM is trained, we map all images onto the final SOM and plot a density map in order to see the neurons which were most frequently selected as the BMU (Figure \ref{fig:bmu_count}). The neuron (0, 6) is selected as the BMU for the highest number of input images, 5047 images which accounts for approximately 2$\%$ of the total input images. Moreover, the neurons in the first row were more likely to be selected as BMU, showing that even for `multi-Gaussian' sources relatively simple morphologies are dominant. Whereas the neurons (3, 1) and (4, 2) which we label as Triple sources (see Section \ref{annotation} for a description of the labels) are selected as BMU the least number of times, for 548 and 621 input images respectively. Generally, we see that the neurons towards the edges, especially in the bottom half of the grid, were chosen to be the BMU much less frequently than those near the top half. We also see that the impact of the circular region over which the Euclidean distances are calculated are more visible for some neurons than others and can be seen in the SOM grid, for example neuron (9, 9) at the bottom right. We also note that structures outside this region are essentially noise and potentially do not carry any real meaning.

Another important property of a SOM to consider is coherence which can be a useful tool for morphological studies such as this. The coherence gives us the total number of times where the neuron which was the next best match for an input image (i.e. had the second lowest value of Euclidean distance between it and the input image) was neighbouring (both next to or diagonally) the neuron chosen as BMU for the same image. A high coherence value indicates that neighbouring neurons are similar to each other and the underlying structures of the SOM are well-organised and represented in the SOM grid. The coherence value in our SOM is 229678 which means that for 91.4$\%$ of the input images, their second closest best-matching neuron is adjacent to the best-matching neuron in the grid. This shows that the different neighbourhoods in our SOM grid are well-established and so the regions in and of themselves are also a useful metric of morphology and not just the precise coordinate. We see this in our trained SOM (Figure \ref{fig:som_final}) where similar representative images or morphologies are closer to each other and can be grouped into similar morphological classes. 

\begin{figure}[h!]
    \centering
    \includegraphics[width=0.9\columnwidth]{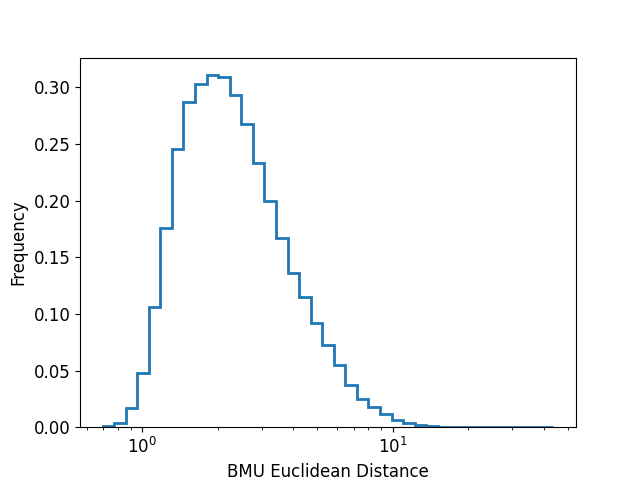}
    \caption{The distribution of the Euclidean distance between input images and their corresponding BMU neuron.}
    \label{fig:neuron_distance_log}
\end{figure}

We plot the distribution of the Euclidean distance between all the input images and their corresponding BMU (Figure \ref{fig:neuron_distance_log}; distributions of the Euclidean distances for each individual neurons in the SOM can be found in the \ref{appendix1} Figures \ref{fig:top_left}, \ref{fig:top_right}, \ref{fig:bottom_left}, and \ref{fig:bottom_right}). The distance in this paper ranged from 0.69 to 43.82, where larger Euclidean distance values indicate there are larger differences between an image and the neuron. It should be noted that the Euclidean distance values generally depend on the input images used to train the SOM as well as the specific training configurations of the system and so is therefore unique to each SOM. A visual inspection can be done to establish an approximate distance threshold at which point the input image starts to no longer resemble its associated BMU. 

\begin{table}[h]
  \centering
  \begin{tabular}{c|c|c|c}
    \textbf{Intervals} & \textbf{Euclidean Distance} & \textbf{Total Sources} & \textbf{Sample Sources} \\
    \midrule
    1 & 0.69 $\leq$ ED < 2.74 & 114392 & 338 \\
    2 & 2.74 $\leq$ ED < 4.78 & 83327 & 288 \\
    3 & 4.78 $\leq$ ED < 6.83 & 33501 & 183 \\
    4 & 6.83 $\leq$ ED < 8.87 & 12139 & 110 \\
    5 & 8.87 $\leq$ ED < 10.91 & 4648 & 68 \\
    6 & 10.91 $\leq$ ED < 12.96 & 1845 & 42 \\
    7 & 12.96 $\leq$ ED < 15.00 & 739 & 27 \\
    8 & 15.00 $\leq$ ED $\leq$ 43.82 & 668 & 25 \\
  \end{tabular}
  \caption{Intervals based on Euclidean distance between randomly chosen input images and their BMU.}
  \label{tab:intervals_sources}
\end{table}

\begin{figure*}[hp!]
    \centering
    %Upper panel (a)
    \begin{subfigure}[b]{1.0\textwidth}
        \includegraphics[width=\textwidth]{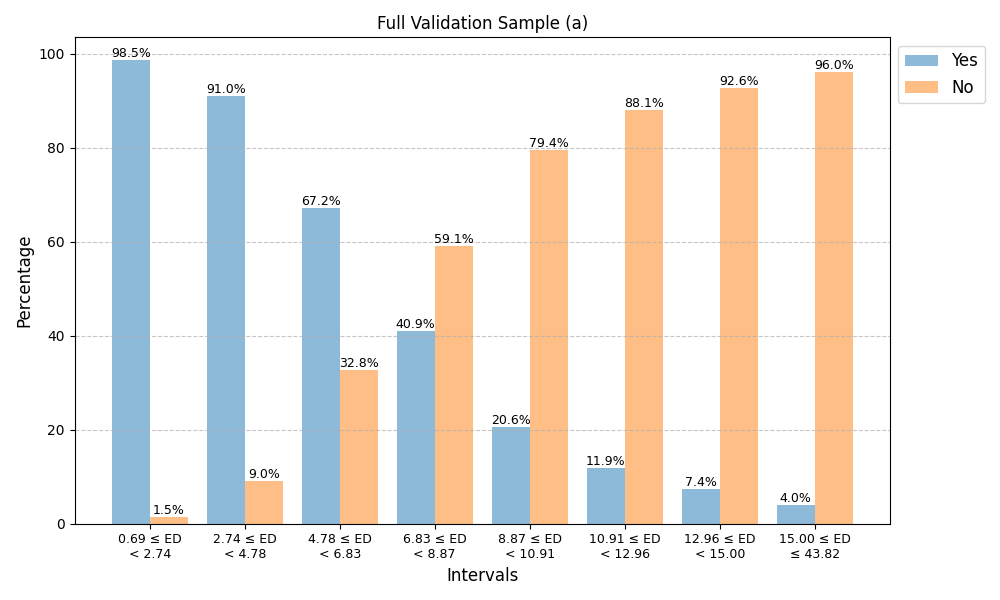}
    \end{subfigure}
    %Lower panels (b-e)
    \begin{subfigure}[b]{0.495\textwidth}
        \centering
        \includegraphics[width=\textwidth]{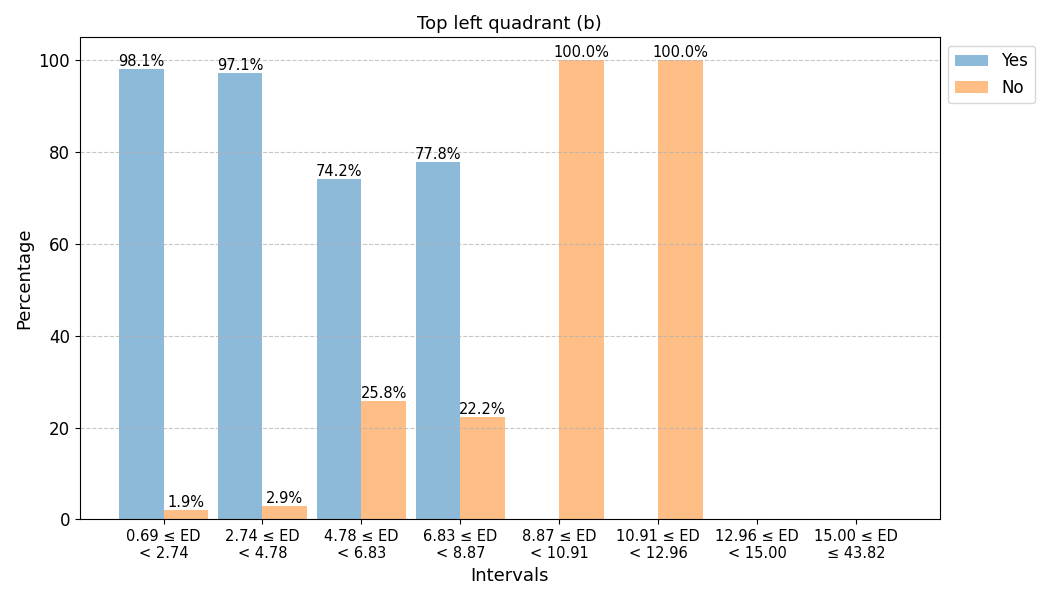}
    \end{subfigure}
    \begin{subfigure}[b]{0.495\textwidth}
        \includegraphics[width=\textwidth]{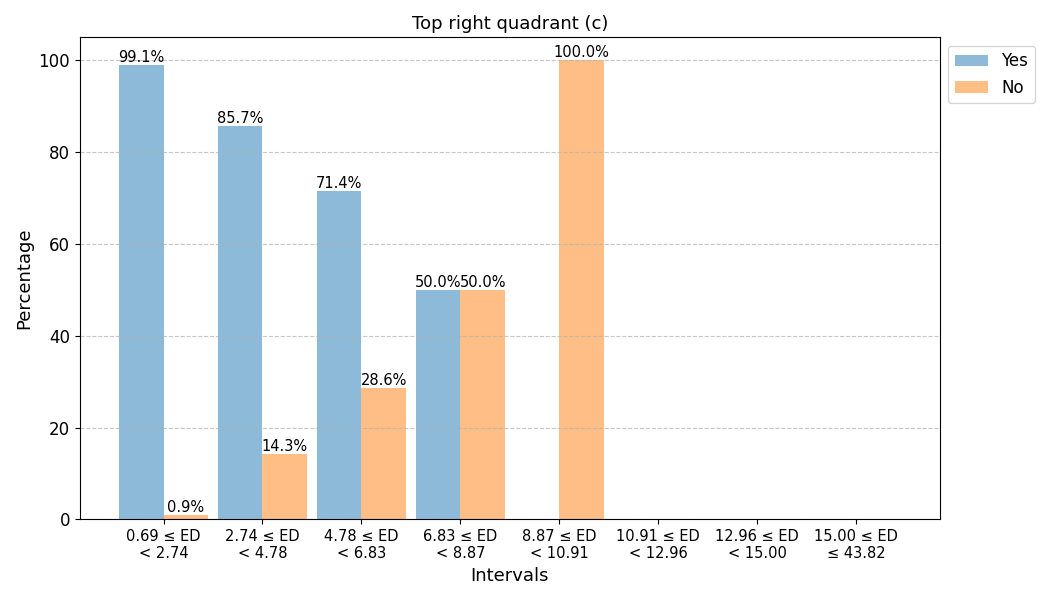}
    \end{subfigure}
    \begin{subfigure}[b]{0.495\textwidth}
        \includegraphics[width=\textwidth]{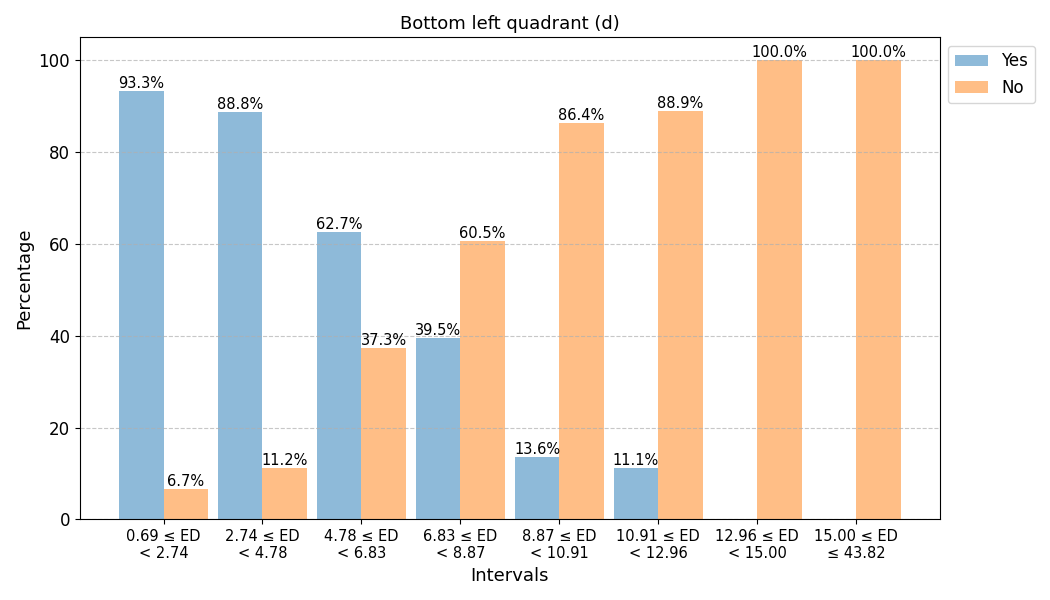}
    \end{subfigure}
    \begin{subfigure}[b]{0.495\textwidth}
        \includegraphics[width=\textwidth]{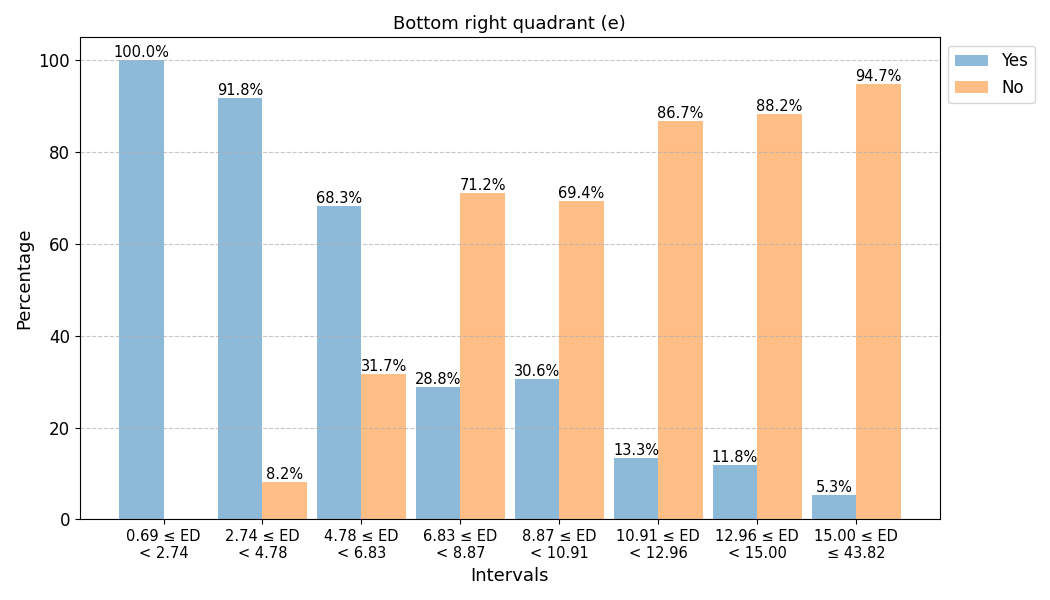}
    \end{subfigure}

    \caption{Upper panel (a): A manual validation of the match between original input images in the full validation sample and their corresponding BMU, where the sample is divided into smaller intervals on the Euclidean distance (Table \ref{tab:intervals_sources}). 
    Lower panels (b-e): Distribution of the `Yes' and `No' matches from the validation scheme above split into the SOM quadrants: top left quadrant (b), top right quadrant (c), bottom left quadrant (d), and bottom right quadrant (e).}
    \label{fig:validation}
\end{figure*}

\begin{figure*}
    \centering
    \includegraphics[width=0.92\textwidth]{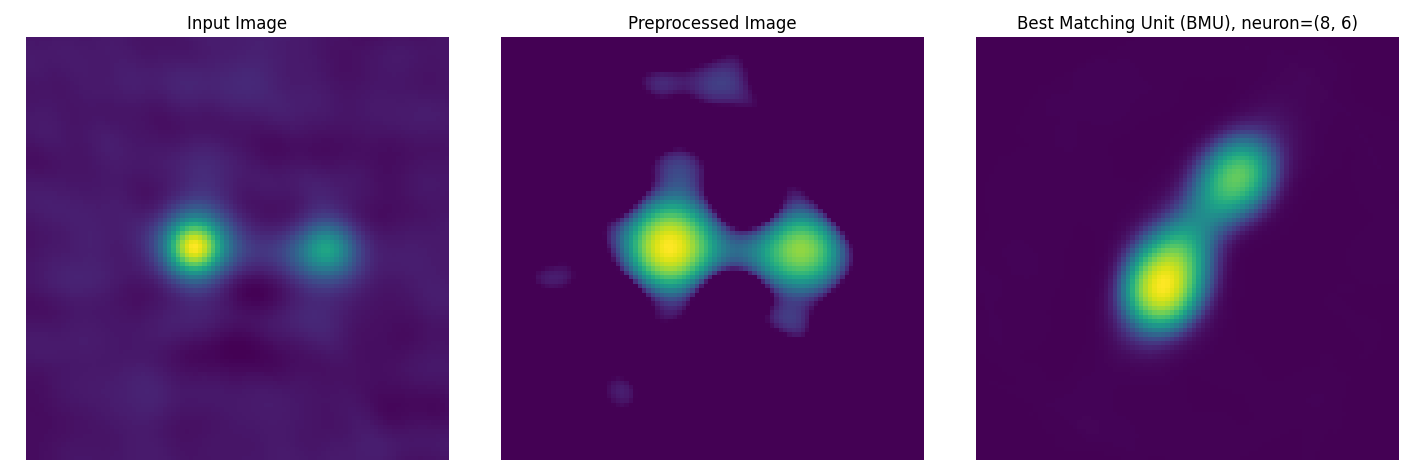}
    \hfill
    \includegraphics[width=0.92\textwidth]{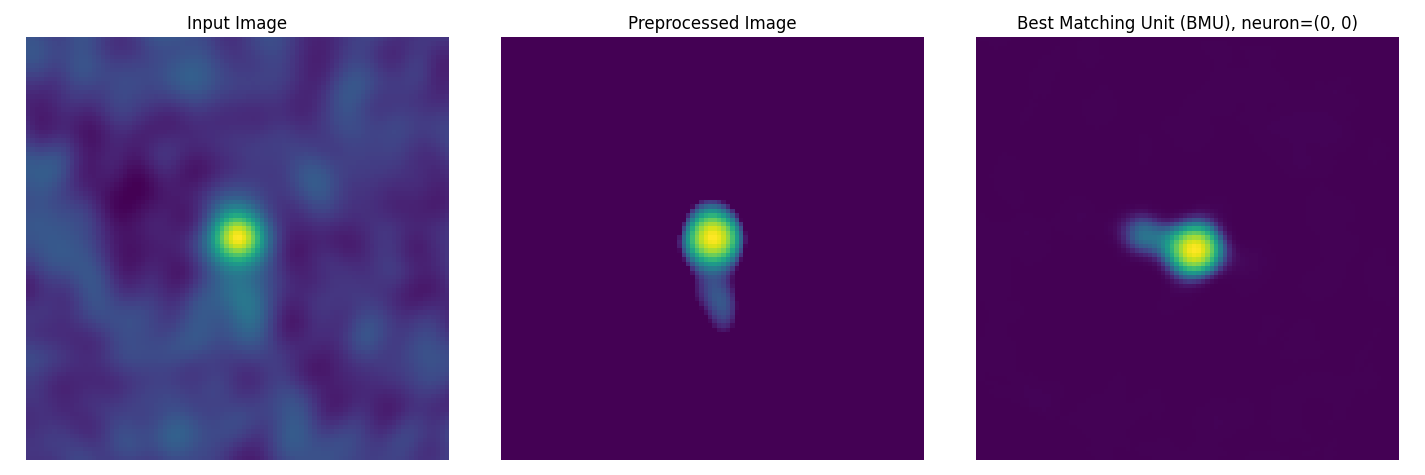}
    \caption{Examples of a `Yes' match for an Input Image, and its corresponding Preprocessed Image and Best-Matching Unit (BMU).}
    \label{fig:som_match_yes}
\end{figure*}

\begin{figure*}
    \centering
    \includegraphics[width=0.92\textwidth]{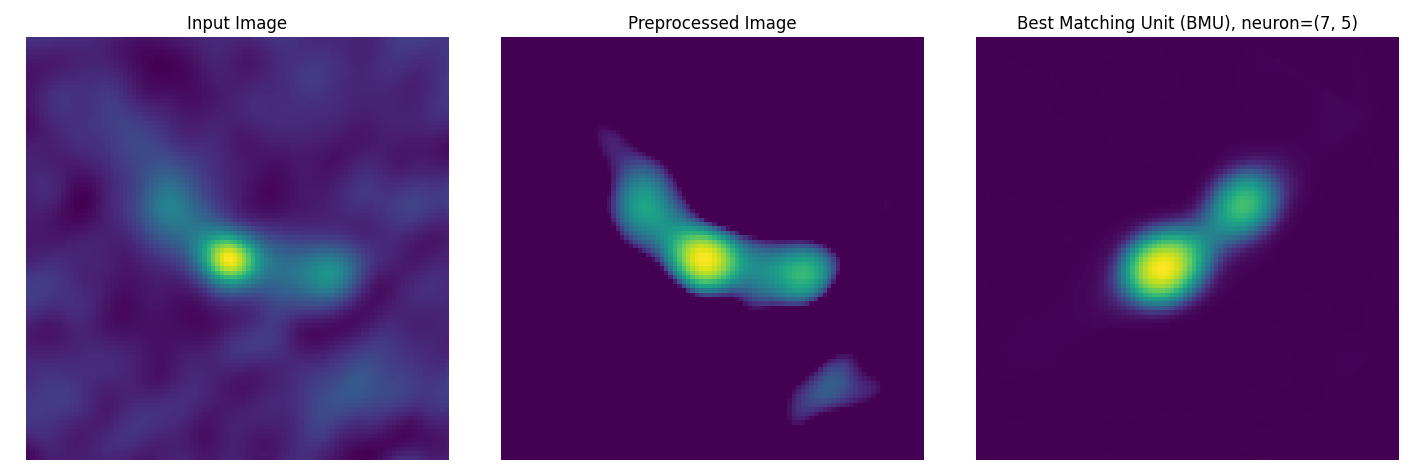}
    \hfill
    \includegraphics[width=0.92\textwidth]{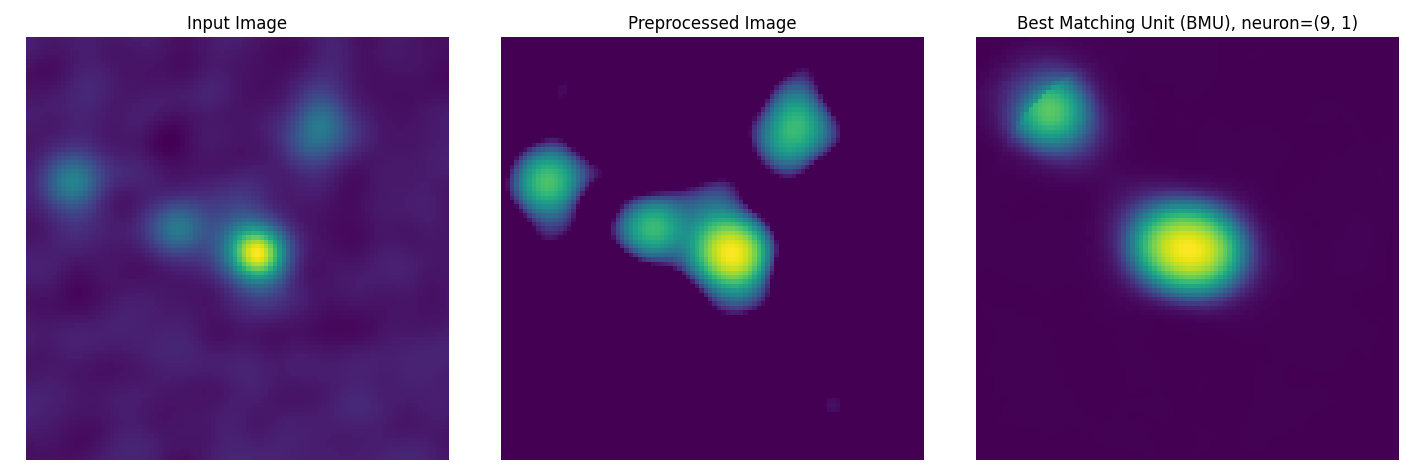}
    \caption{Examples of a `No' match for an Input Image, and its corresponding Preprocessed Image and Best-Matching Unit (BMU).}
    \label{fig:som_match_no}
\end{figure*}

To perform a quantitative validation of the similarity between input image cutout and the BMU we create a smaller validation sample by dividing the range of Euclidean distances into 8 intervals and selecting random sources from each (Table \ref{tab:intervals_sources}). The first 7 intervals comprised of distances in the range 0.69 to 15.0 as 99.7$\%$ of our dataset have distances in that range, and the 8th interval consisted of those with distances 15.0 to 43.82. Next, we take the $\sqrt{n}$ of the number of sources in each interval to create the validation sample so a manual validation of the matches can be performed in a more efficient and less time-consuming manner. Prior to the validation, we inspect multiple random input images and their BMU to review the criteria of a match so as to avoid bias and ensure consistent labelling. A visual inspection is subsequently done on the validation sample by looking at each of the original input images in the sample and their BMU to see if they matched or mostly matched (rotating and flipping the input images to match the BMUs if needed), such that the input image can be reasonably believed to have contributed to the BMU neuron which is an aggregate of all input images for which it is chosen as the BMU. These matches were designated as `Yes' with the others being assigned as `No' (the upper panel a in Figure \ref{fig:validation}). While this is a subjective match based upon our visual inspection, it provides a good baseline for reliability in the similarity matches (see Figures \ref{fig:som_match_yes} and \ref{fig:som_match_no} for examples of how the similarity was judged). 

As expected, the number of `Yes' matches between the input and the BMU decrease with Euclidean distances. In Intervals 1 and 2 (which cover the distance 0.69 - 4.78 overall and comprise of 78.7$\%$ of our dataset), the number of `Yes' matches are 98.5$\%$ and 91.0$\%$ respectively and can be considered to be the most reliable. In Interval 3, the percentage of `Yes' matches (67.2$\%$) have decreased from the previous intervals, but is still considerably higher than the `No' matches. In Interval 4 and beyond (Euclidean distance larger than 6.83), the number of `Yes' matches continue to decrease and are significantly less than their `No' counterparts. As a result, their morphological labels become less reliable, with Intervals 7 and 8 being the least reliable given that more than 90$\%$ of their input images do not visually match their BMU. This provides us with a potential second subset of unique and complex galaxies for further study, and can enable us to potentially find rare and unexpected objects within the RACS catalogue. Therefore, for Euclidean distances to the BMU $\lesssim$ 5, which account for approximately 79$\%$ of our sources, the reliability of the morphological labels in our catalogue is estimated to be $>90\%$, and this reliability drops down to less than 70$\%$ at Euclidean distances $\gtrsim$ 7. We also qualitatively group the validation sample into SOM sub-regions and plot the distribution of the `Yes' and `No' matches to see if it varies depending on the region (the lower panels b-e in Figure \ref{fig:validation}). This is done by splitting the full validation sample according to which SOM quadrant (see Figure \ref{fig:som_final} for the four quadrants marked in red) the BMU of the validation sources are located in: top left (235 sources), top right (285 sources), bottom left (264 sources), and bottom right (297 sources). It should be noted that since the sampling for the validation subset was done randomly we do not have equal numbers of each quadrant present. However, the general trend of a higher number of `Yes' matches than `No' at lower Euclidean distances, particularly for the first three intervals, is still present. In addition, for the top left (panel a) and top right (panel b) quadrants we see that there are fewer sources with high Euclidean distances that fall within the intervals on the right hand side unlike the bottom quadrants (panels c and d). This indicates that the sources in the top quadrants skew towards lower Euclidean distances. 

\begin{figure*}[h!]
    \centering
    \includegraphics[width=\columnwidth]{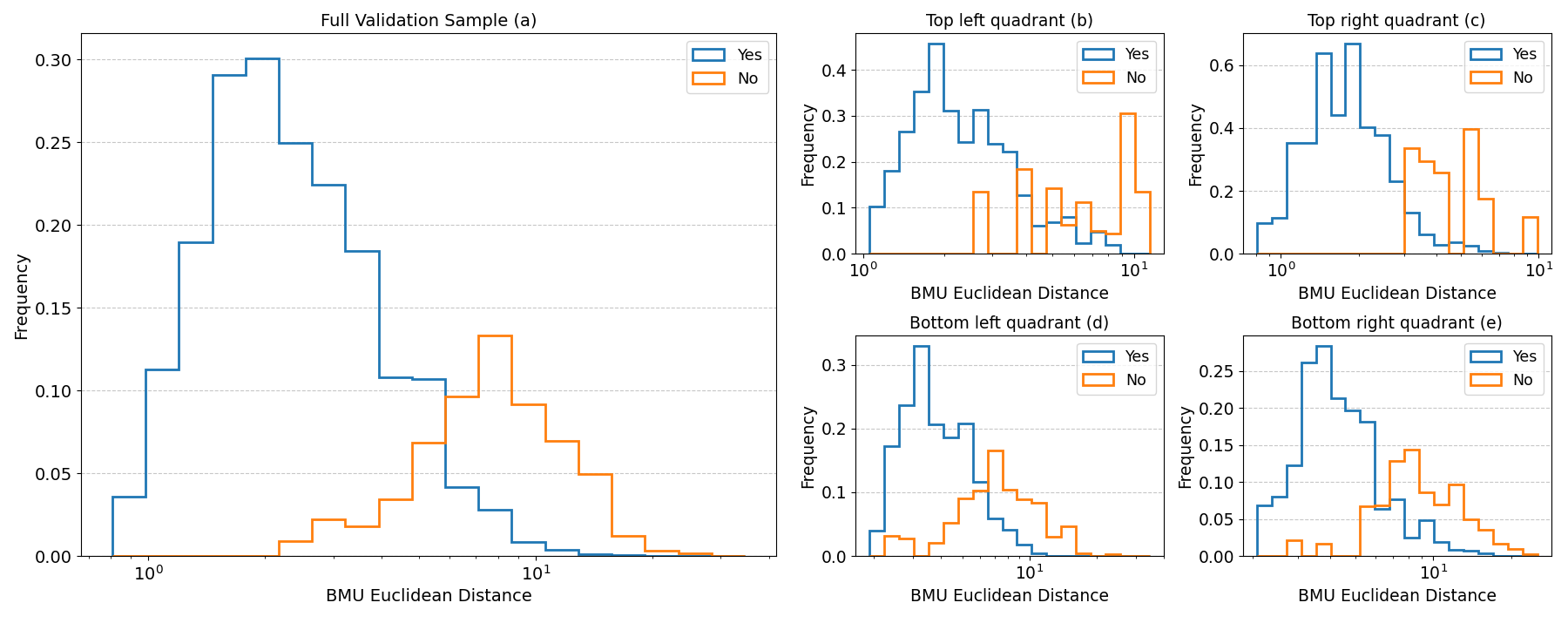}
    \caption{The distribution of the Euclidean distance between input images and their corresponding BMU for the `Yes' and `No' matches in the validation sample (a). The distance distributions for the sources in the validation sample grouped into SOM regions: Top left quadrant (b), top right quadrant (c), bottom left quadrant (d), and bottom right quadrant (e).}
    \label{fig:yes_no_euclidean}
\end{figure*}

Subsequently, we plot the distribution of the Euclidean distances between all the input images and their corresponding BMU in the validation sample for both `Yes' and `No' matches (panel a in Figure \ref{fig:yes_no_euclidean}), as well as the split into the SOM quadrants (panels b to e). For the full validation sample we see that the distribution for `Yes' matches peaks at $\sim$2, and the skews heavily towards the left-hand side of the graph at lower Euclidean distances with the majority of the values falling within distances of $\sim$5. Whereas, the distribution for `No' matches span a wider range of Euclidean distances, and when compared to the `Yes' matches it skews more towards the higher end of Euclidean distances. This indicates that at higher Euclidean distances we expect the fraction of input images which are similar to their BMU to start decreasing. This trend generally holds in the four SOM quadrants with `Yes' matches skewing mostly towards lower Euclidean distances, and the `No' matches leaning more towards higher distances in comparison (panels b-e in Figure \ref{fig:yes_no_euclidean}). We can also see that for the top left and top right quadrants there are fewer `No' matches on the left-hand side which is in contrast to the bottom left and bottom right quadrants where the distributions vary over a wider range of distances. These results can be attributed to the neurons in the top SOM quadrants generally being dominated by relatively simple and smaller structures, whereas the neurons at the bottom quadrants have larger or more extended sources. These larger and more extended sources are more likely to have higher levels of background and noise when compared to the simpler sources even following image preprocessing. In addition, their extended sizes could prevent them from being fully captured by the current cutout size of \ang{;4;}. As a result, they might not be modelled as well as simpler sources during the SOM training. As such, we would expect the Euclidean distances between the top quadrant neurons and their input images to be comparatively lower than for the bottom quadrants (see Figures \ref{fig:top_left}, \ref{fig:top_right}, \ref{fig:bottom_left}, \ref{fig:bottom_right} in \ref{appendix1} for the distributions of the Euclidean distances split by individual neurons in each quadrant for the whole dataset).

\subsection{Annotating the SOM}
\label{annotation}

Once a reliability threshold for the similarity between an input image and its BMU had been established, the next stage is to manually label or tag each of the 100 neurons based on their shown morphology. We have decided on 6 tags which broadly encompasses the morphologies seen in the SOM grid: 
\begin{itemize}
    \item \textbf{C} (Compact) sources without any significant features other than the central core and are circular or nearly circular.
    \item \textbf{EC} (Extended Compact) sources which are compact sources with either a bright central core and some extended structures, such as an elongated compact core, a tail or additional neighbouring components.
    \item \textbf{CD} (Connected Double) sources which comprise of two distinguishable lobes of comparable sizes or brightness which are either connected or the angular distance between them is relatively minimal.
    \item \textbf{SD} (Split Double) sources which comprise of two distinguishable lobes of comparable sizes or brightness with a clear angular separation where the separation is relatively large.
    \item \textbf{T} (Triple) sources which comprise of three distinguishable lobes of relatively comparable sizes or brightness.
    \item \textbf{U/A} (Uncertain/Ambiguous) sources which contains characteristics that might be present in more than one of the previous labels and it is not clear which label would be the best fit.
\end{itemize}

\begin{figure*}[p]
    \centering
        \includegraphics[width=1.0\textwidth]{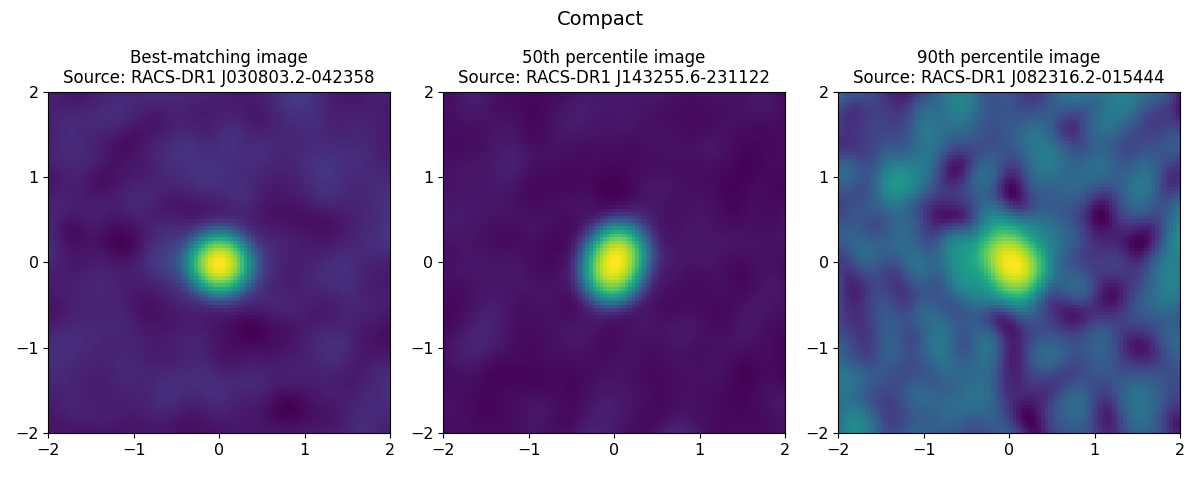}
        \vspace{1em}
        \includegraphics[width=1.0\textwidth]{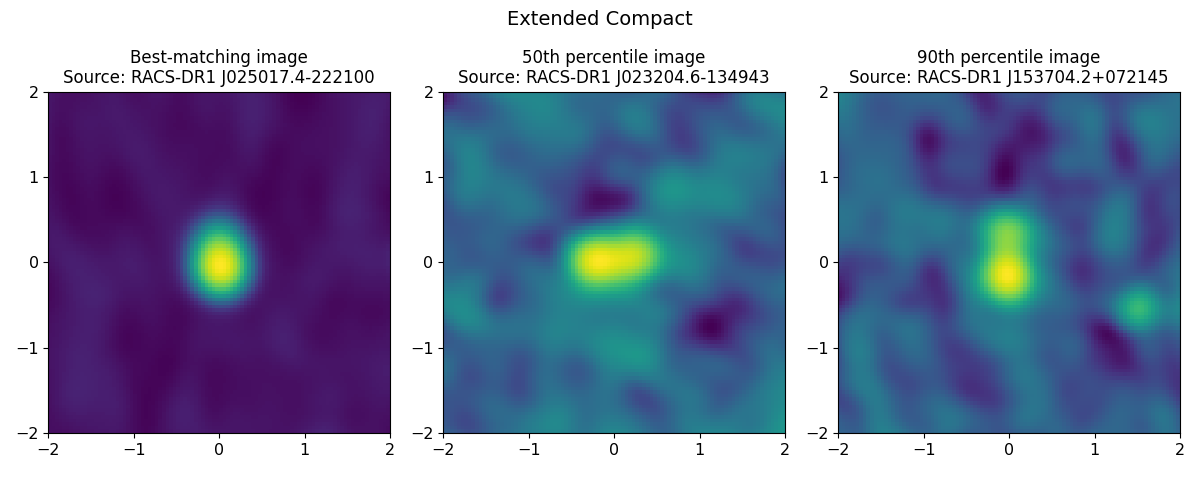} 
        \vspace{1em}
        \includegraphics[width=1.0\textwidth]{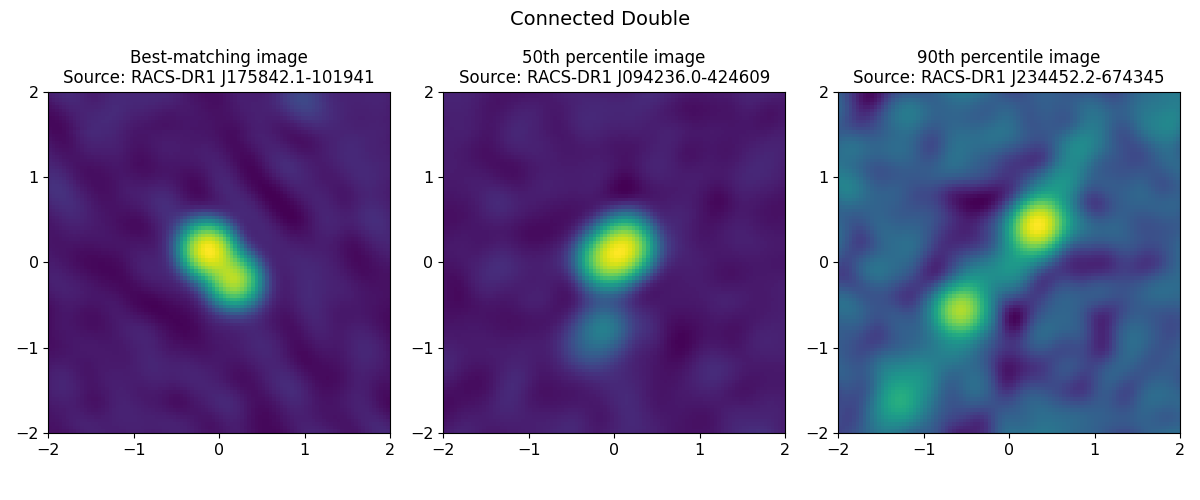} 
    \caption{The best-matching, 50th percentile and 90th percentile images for the labels Compact (C), Extended Compact (EC) and Connected Double (CD).}
    \label{fig:validation_examples}
\end{figure*}

\begin{figure*}[p]\ContinuedFloat
     \centering
        \includegraphics[width=1.0\textwidth]{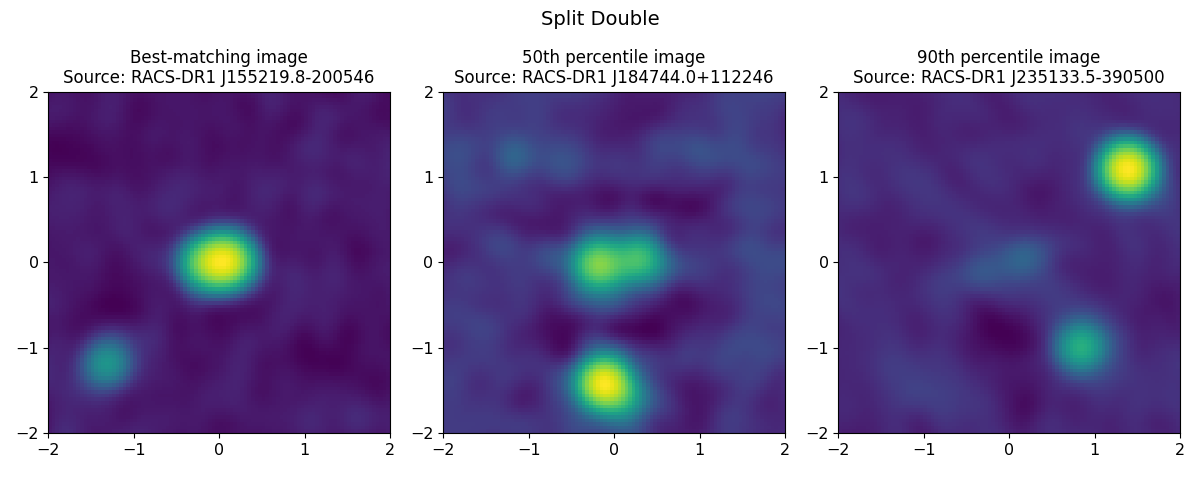}
        \vspace{1em}
        \includegraphics[width=1.0\textwidth]{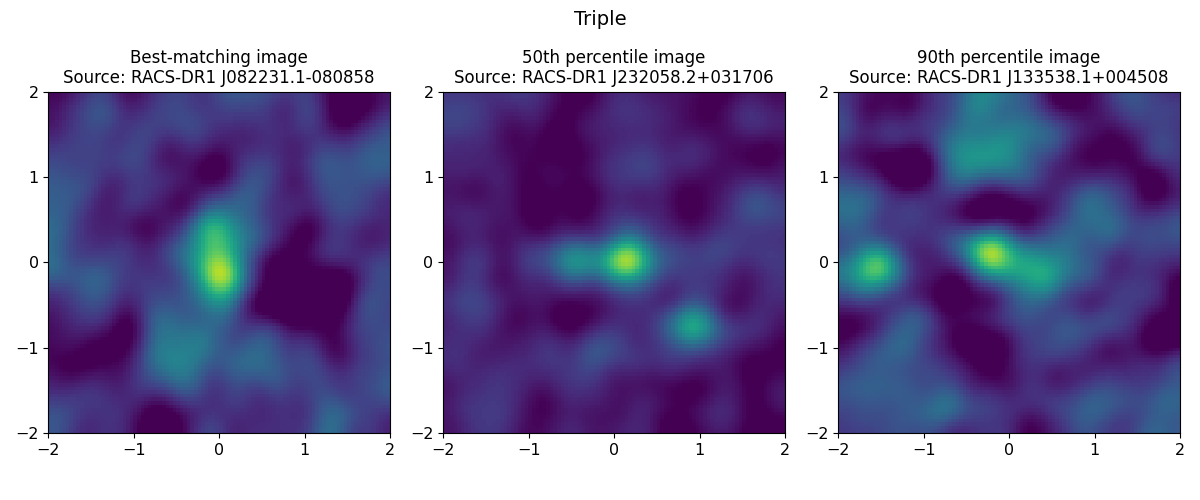} 
        \vspace{1em}
        \includegraphics[width=1.0\textwidth]{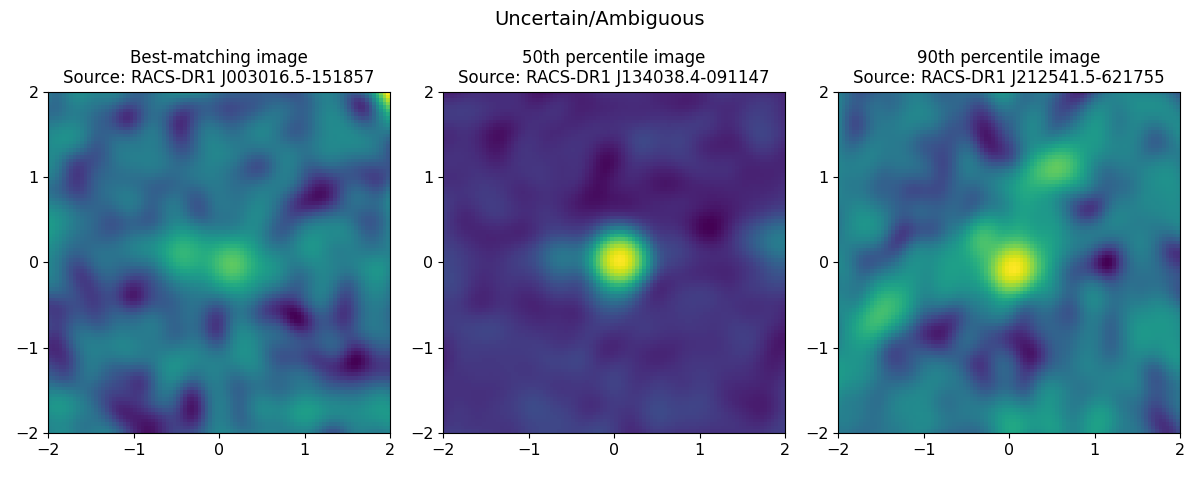} 
    \caption{The best-matching, 50th percentile and 90th percentile images for the labels Split Double (SD), Triple (T) and Uncertain/Ambiguous (U/A).} 
    \label{fig:validation_examples_2}
\end{figure*}

\begin{table*}[!ht]
    \centering
    \begin{tabular}{c|c|c|c|c|c|c|c|c|c|c}
    \textbf{Morphological Label} & \textbf{Neurons} & \textbf{Total Sources} & \textbf{98.5\%} & \textbf{91.0\%} & \textbf{67.2\%} & \textbf{40.9\%} & \textbf{20.6\%} & \textbf{11.9\%} & \textbf{7.4\%} & \textbf{4.0\%} \\
    \midrule
    Compact & 6 & 26217 & 23964 & 2225 & 28 & 0 & 0 & 0 & 0 & 0 \\
    Extended Compact & 25 & 81396 & 59920 & 18964 & 2250 & 238 & 23 & 1 & 0 & 0 \\
    Connected Double & 48 & 106671 & 29181 & 48612 & 16915 & 6745 & 2866 & 1273 & 529 & 550 \\
    Split Double & 14 & 25379 & 18 & 6644 & 11345 & 4753 & 1724 & 567 & 210 & 118 \\
    Triple & 2 & 1169 & 2 & 317 & 669 & 165 & 13 & 3 & 0 & 0 \\
    Uncertain/Ambiguous & 5 & 10427 & 1307 & 6565 & 2294 & 238 & 22 & 1 & 0 & 0 \\
    \end{tabular}
    \caption{Summary of the classification of the morphological labels. From left to right we give the morphological label, the number of neurons which were assigned said label, the total number of sources in the RACS catalogue once the labels were transferred, and the split of the sources into each reliability percentage from the validation process based on Euclidean distances.}
    \label{tab:summary}
\end{table*}

We tag the individual neurons with the aforementioned labels and transfer these labels to each of the input images from their BMU (see Figure \ref{fig:validation_examples} to see examples of the best-matching, 50th percentile and 90th percentile input images for each of the labels). Once the labels have been transferred, we can combine them with the reliability percentage from the validation scheme. Table \ref{tab:summary} gives a summary of the classification of the morphological labels along with the number of neurons which have been assigned these labels, the total number of sources in the RACS catalogue after label transfer, and the split of these sources into the reliability percentages from the validation process: 98.5\%, 91.0\%, 67.2\%, 40.9\%, 20.6\%, 11.9\%, 7.4\% and 4.0\%. The labels `Compact' and `Extended Compact' can be considered the most reliable since a greater fraction of their total sources (91.4\% and 73.6\% respectively) have a reliability of $\gtrsim$ 98.5\%. This is to be expected given their relatively simple morphologies. We also see that the majority of neurons assigned these labels are located in the top left and top right quadrants of the SOM which generally has lower Euclidean distances both in the validation sample (Figure \ref{fig:yes_no_euclidean}) and in the overall SOM (Figures \ref{fig:top_left}, \ref{fig:top_right}, \ref{fig:bottom_left}, \ref{fig:bottom_right} in \ref{appendix1}). Approximately 73\% of `Connected Double' sources fall within the reliability threshold of 91.0\% or higher which demonstrates an overall good match. For `Split Double' and `Triple' sources, the biggest fraction of sources have around 67.2\% reliability match and so these labels can be considered moderately reliable. It should be noted that the label `Triple' has the fewest number of sources, approximately 0.45\% of the total sources, and the two neurons assigned this label are also least commonly chosen as BMU (Figure \ref{fig:bmu_count}). As such, while the reliability percentage for this label is relatively moderate, the limited number of sources does limit the extent to which the label can be utilised. Majority of the `Uncertain/Ambiguous' sources have a high reliability match of $\gtrsim$ 91.0\% which indicates that the associated neurons capture the ambiguous nature of their morphologies well. We can further assess the reliability of the transferred morphological label for each source by also considering the morphological labels of its next best-matching neurons to see if they are consistent. We find that for 75.55\% of sources, the labels of its second best-matching neuron matches the label from its BMU. For third and fourth best-matching neurons the percentage of matches with the BMU label decreases to 71.83\% and 68.98\% respectively. This aligns with our expectations as the similarity between the source and the neuron will diverge with increasing Euclidean distance. Therefore, for the majority of sources the transferred morphological labels from the BMU are consistent with the labels of its next best-matching neurons, especially the second best-matching neuron.

\begin{figure*}
    \centering
    \includegraphics[width=0.275\textwidth]{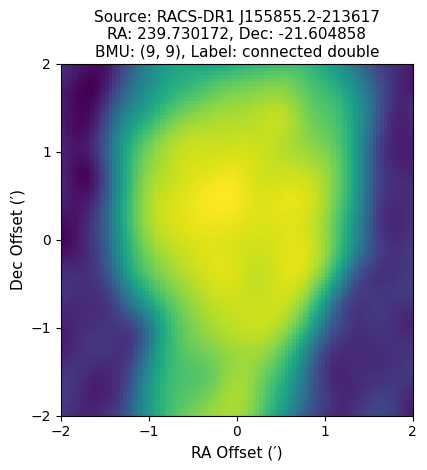}
    \hspace{0.75cm} 
    \includegraphics[width=0.275\textwidth]{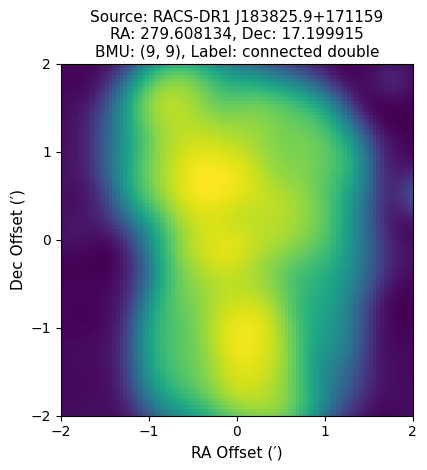}
    \hspace{0.75cm} 
    \includegraphics[width=0.275\textwidth]{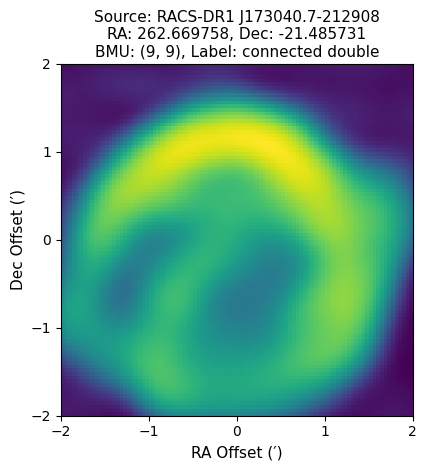}
    %\hspace{0.75cm} 
    \includegraphics[width=0.275\textwidth]{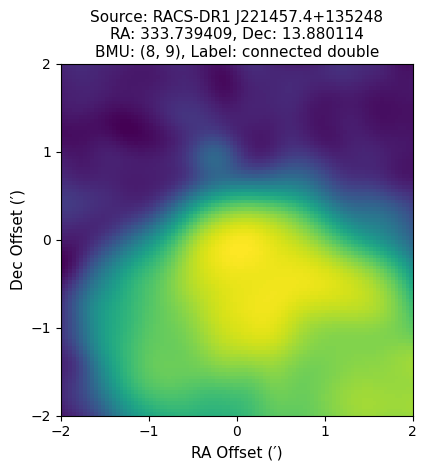}
    \hspace{0.75cm} 
    \includegraphics[width=0.275\textwidth]{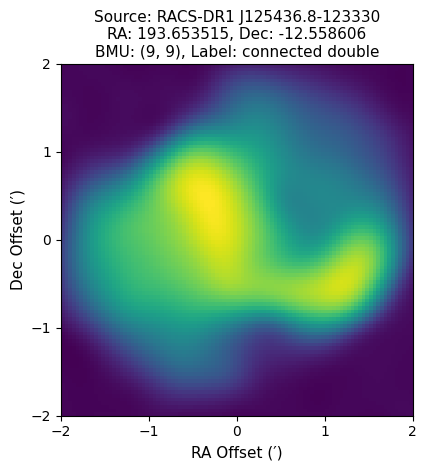}
    \hspace{0.75cm} 
    \includegraphics[width=0.275\textwidth]{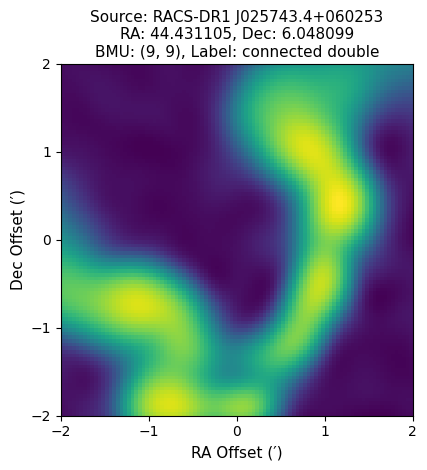}
    %\hspace{0.75cm} 
    \includegraphics[width=0.275\textwidth]{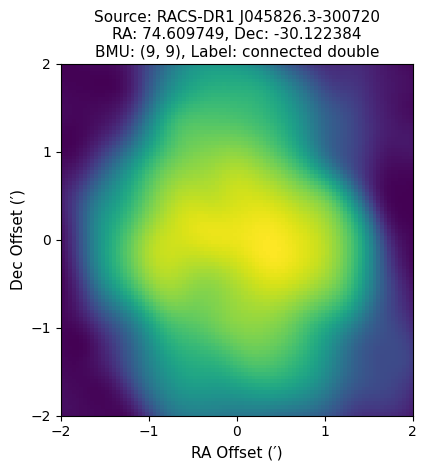}
    \hspace{0.75cm} 
    \includegraphics[width=0.275\textwidth]{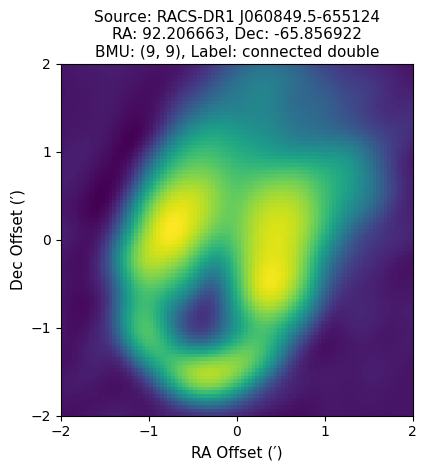}
    \hspace{0.75cm} 
    \includegraphics[width=0.275\textwidth]{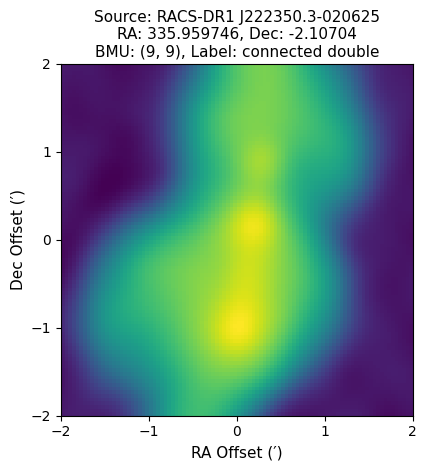}
    %\hspace{0.75cm} 
    \includegraphics[width=0.275\textwidth]{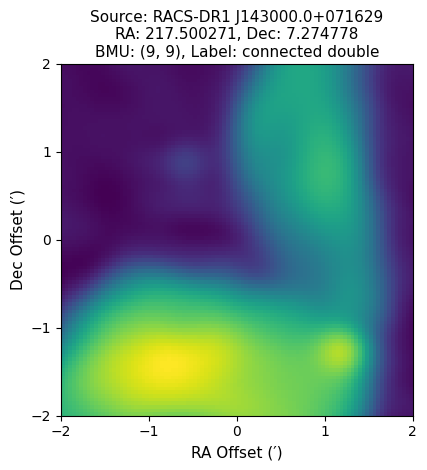}
    \hspace{0.75cm} 
    \includegraphics[width=0.275\textwidth]{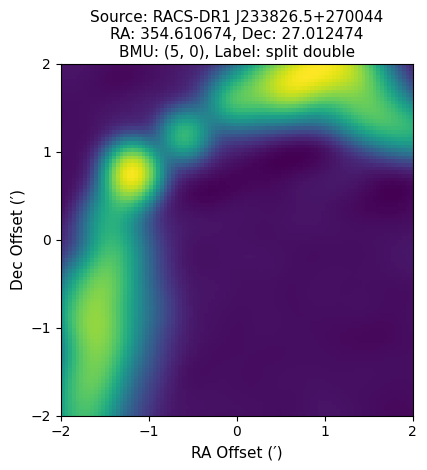}
    \hspace{0.75cm} 
    \includegraphics[width=0.275\textwidth]{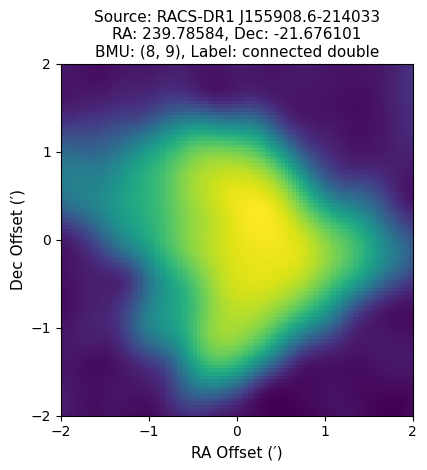}
    \caption{The 12 sources with the largest Euclidean distances between their input images and BMU. For each source we give its source name, RA, Dec, BMU in the SOM grid and its morphological label after label transfer. The distance for these sources range from 34.36 to 43.82 and they all have a reliability percentage of around 4.0\% based on the validation scheme.}
    \label{fig:large_ed}
\end{figure*}

It is important to note that these labels are not universally agreed upon labels but are instead subjective and based upon our visual inspections of the neurons on this specific trained SOM. As such, they are not transferable to other SOMs even if they are trained on the same data. However, the labels once transferred to our value-added catalogue of RACS sources will help us quickly distinguish the broad morphological features present within the dataset (\citealp{Rudnick2021}, \citealp{Bowles2023}). Moreover, it can help us quickly identify atypical and rare sources. In Figure \ref{fig:large_ed} we show the 12 sources with the largest Euclidean distances between their input images and their BMU. These sources have relatively unusual and interesting morphologies, and their high BMU Euclidean distances is due to them not being very well represented by the neurons. These sources have a reliability percentage of around 4.0\% from our validation scheme, which further indicates that their SOM-derived morphological labels are not very dependable. A more thorough study of these sources will be done in the next paper.

%%%%%%%%%%%%%%%%%%%%%%%%%%%%%%%%%%%%%%%%%%%%%

\subsection{Catalogue of Complex Sources}
\label{catalogue}

For each source in our dataset, we add the position of its BMU in the SOM grid, the Euclidean distance between it and the BMU, the transferred morphological label based on the visual inspection of the individual neurons, and the reliability percentage based on the Euclidean distance to create our final catalogue of complex sources. The catalogue contains the following columns: \texttt{source\_name}, \texttt{source\_id}, \texttt{ra}, and \texttt{dec} from the RACS catalogue, \texttt{bmu} which gives the position of the best-matching neuron in the SOM grid, \texttt{euclidean\_distance} which gives the Euclidean distance between the BMU and the input image, the \texttt{morphological\_label} based on observed morphological features present based on the visual inspections, \texttt{match\_percent} which gives the reliability percentage, i.e. the percentage of sample input images which matched with its BMU based on the Euclidean distance (see Table \ref{tab:catalogue_description} for more details on the columns). The first 30 rows of the catalogue are shown in Table \ref{tab:catalogue_example}. The full catalogue produced in this paper will be made available in CDS VizieR (\citealp{VizieR}) and other key databases after publication.

\begin{table}[!t]
  \centering
  \begin{tabular}{c|p{5cm}}
    \textbf{Catalogue column} & \textbf{Description} \\
    \midrule
    \texttt{source\_name} & Name of the source as given in the RACS catalogue which follows the IAU convention JHHMMSS.S±DDMMSS with the prefix RACS-DR1.\\
    \texttt{source\_id} & ID of the source as given in the RACS catalogue which is the RACS tile ID along with the Src\_ID generated by \texttt{PyBDSF}.\\
    \texttt{ra} & Right Ascension coordinate of the source (in degrees). \\
    \texttt{dec} & Declination coordinate of the source (in degrees). \\
    \texttt{bmu} & The position of the Best-Matching Unit (BMU), i.e. the neuron which best-matched the input image, in the SOM grid (Figure \ref{fig:som_final}). The neuron coordinates are in the form (y, x).\\
    \texttt{euclidean\_distance} & The Euclidean distance between the source and BMU in the SOM grid.\\
    \texttt{morphological\_label} & The SOM-derived morphological label (see \ref{annotation} for more information on the labels used).\\
    \texttt{match\_percent} & The reliability percentage which gives the percentage of input images which matched with its BMU based on visual inspection of a smaller validation sample.\\
  \end{tabular}
  \caption{Description of the columns in the catalogue created in this paper.}
  \label{tab:catalogue_description}
\end{table}

\begin{table*}
\centering
    \begin{tabular}{llrrlrll}
    \toprule
    source\_name &          source\_id &       ra &       dec &    bmu &  euclidean\_distance & morphological\_label & match\_percent \\
    \midrule
    RACS-DR1 J001232.8+135445 & RACS\_0000+12A\_1102 & 3.136777 & 13.912659 & (1, 3) &            1.613091 &    extended compact &         98.5\% \\
    RACS-DR1 J001218.2+120733 & RACS\_0000+12A\_1115 & 3.076231 & 12.125925 & (9, 6) &            5.108846 &    connected double &         67.2\% \\
    RACS-DR1 J001217.3+140104 & RACS\_0000+12A\_1122 & 3.072128 & 14.017922 & (4, 9) &            3.465037 &    extended compact &         91.0\% \\
    RACS-DR1 J001213.7+134434 & RACS\_0000+12A\_1137 & 3.057311 & 13.742932 & (2, 1) &            4.860659 &    connected double &         67.2\% \\
    RACS-DR1 J001159.7+111637 & RACS\_0000+12A\_1141 & 2.998874 & 11.277090 & (9, 6) &            4.630018 &    connected double &         91.0\% \\
    RACS-DR1 J001201.7+120116 & RACS\_0000+12A\_1149 & 3.007460 & 12.021187 & (0, 3) &            1.390036 &    extended compact &         98.5\% \\
    RACS-DR1 J001155.1+101847 & RACS\_0000+12A\_1155 & 2.979664 & 10.313124 & (9, 7) &            7.687910 &    connected double &         40.9\% \\
    RACS-DR1 J001152.5+125156 & RACS\_0000+12A\_1174 & 2.968830 & 12.865726 & (5, 8) &            3.627663 &    connected double &         91.0\% \\
    RACS-DR1 J001156.9+135007 & RACS\_0000+12A\_1178 & 2.987112 & 13.835327 & (7, 1) &            4.852356 &        split double &         67.2\% \\
    RACS-DR1 J001140.2+134548 & RACS\_0000+12A\_1210 & 2.917669 & 13.763403 & (7, 3) &            3.632242 &    connected double &         91.0\% \\
    RACS-DR1 J001135.7+124553 & RACS\_0000+12A\_1217 & 2.898931 & 12.764777 & (6, 5) &            4.224261 &    connected double &         91.0\% \\
    RACS-DR1 J001129.2+104835 & RACS\_0000+12A\_1221 & 2.871919 & 10.809794 & (0, 5) &            2.472950 &             compact &         98.5\% \\
    RACS-DR1 J001122.6+101542 & RACS\_0000+12A\_1229 & 2.844409 & 10.261744 & (5, 0) &            5.126052 &        split double &         67.2\% \\
    RACS-DR1 J001131.1+134736 & RACS\_0000+12A\_1230 & 2.879728 & 13.793518 & (1, 8) &            1.701950 &             compact &         98.5\% \\
    RACS-DR1 J001119.8+100738 & RACS\_0000+12A\_1236 & 2.832730 & 10.127255 & (0, 2) &            2.427596 &    extended compact &         98.5\% \\
    RACS-DR1 J001115.9+111800 & RACS\_0000+12A\_1246 & 2.816345 & 11.300248 & (4, 4) &            2.145172 &    connected double &         98.5\% \\
    RACS-DR1 J001115.4+144607 & RACS\_0000+12A\_1265 & 2.814549 & 14.768678 & (0, 6) &            2.099214 &             compact &         98.5\% \\
    RACS-DR1 J001109.8+122838 & RACS\_0000+12A\_1267 & 2.791108 & 12.477418 & (8, 5) &            3.741234 &    connected double &         91.0\% \\
    RACS-DR1 J001108.2+123532 & RACS\_0000+12A\_1272 & 2.784499 & 12.592249 & (0, 6) &            1.701353 &             compact &         98.5\% \\
    RACS-DR1 J001106.3+125027 & RACS\_0000+12A\_1273 & 2.776347 & 12.840854 & (6, 7) &            2.314408 &    connected double &         98.5\% \\
    RACS-DR1 J001030.7+105827 & RACS\_0000+12A\_1336 & 2.628291 & 10.974313 & (1, 8) &            1.747133 &             compact &         98.5\% \\
    RACS-DR1 J001034.6+133848 & RACS\_0000+12A\_1338 & 2.644539 & 13.646904 & (5, 9) &            2.476797 &    connected double &         98.5\% \\
    RACS-DR1 J001023.8+121937 & RACS\_0000+12A\_1363 & 2.599256 & 12.327161 & (2, 3) &            3.860162 &    connected double &         91.0\% \\
    RACS-DR1 J001022.2+134639 & RACS\_0000+12A\_1368 & 2.592654 & 13.777504 & (4, 7) &            3.052129 &    connected double &         91.0\% \\
    RACS-DR1 J001018.2+143337 & RACS\_0000+12A\_1375 & 2.576152 & 14.560517 & (0, 8) &            1.829436 &             compact &         98.5\% \\
    RACS-DR1 J000952.3+124426 & RACS\_0000+12A\_1411 & 2.468288 & 12.740782 & (9, 9) &           15.941885 &    connected double &          4.0\% \\
    RACS-DR1 J000951.3+141738 & RACS\_0000+12A\_1427 & 2.463940 & 14.293970 & (4, 0) &            5.859904 &        split double &         67.2\% \\
    RACS-DR1 J000945.2+141442 & RACS\_0000+12A\_1444 & 2.438447 & 14.245052 & (1, 1) &            2.470556 &    extended compact &         98.5\% \\
    RACS-DR1 J000927.4+095842 & RACS\_0000+12A\_1456 & 2.364251 &  9.978556 & (3, 3) &            1.716938 &    connected double &         98.5\% \\
    RACS-DR1 J000933.5+144146 & RACS\_0000+12A\_1465 & 2.389622 & 14.696120 & (1, 2) &            1.801990 &    extended compact &         98.5\% \\
    \bottomrule
    \caption{The first 30 rows from the final catalogue of complex sources created using the SOM.}
    \label{tab:catalogue_example}
    \end{tabular}
\end{table*}

%%%%%%%%%%%%%%%%%%%%%%%%%%%%%%%%%%%%%%%%%%%%%

\section{Conclusions}
\label{Conclusions}

Next-generation surveys are expected to identify vast number of sources, and as a result will require novel methods of cross-identification. Machine learning methods, especially SOMs, can be used to address the problem of finding complex radio sources in the large dataset provided by SKA pathfinders, such as RACS. In order to do so, we build and train a SOM on sources with multi-Gaussian components from the RACS-Low catalogue. Once the SOM is trained, each input image has a neuron which has been assigned as its best representative or BMU. We label the neurons based on observable morphological structures and then transfer these labels back to the sources from their BMU. This yields a catalogue of complex radio sources, which can be used for further studies. We visually inspect a smaller subset of input images and their BMU to determine a reliability threshold for the similarity metric, which in this case is a modified Euclidean distance. We find that for Euclidean distances of less than 2.74 there is around a 98.5$\%$ chance that a randomly chosen input image will match its BMU, but this percentage decreases with Euclidean distance as expected. This, however, gives us the opportunity to study the most unusual and rare objects present in the data by filtering the catalogue to identify sources with high BMU Euclidean distances, and this will be the topic of our next paper. 

The catalogue created consists of 251,259 objects from RACS-Low and has additional columns added which include: the best-matched neuron or BMU, the Euclidean distance between the input image and its BMU, the morphology label based on its BMU as well as a general confidence level calculated through visual inspections. 

%%%%%%%%%%%%%%%%%%%%%%%%%%%%%%%%%%%%%%%%%%%%%

\begin{acknowledgement}

Y.A.G. is supported by the US National Science Foundation (NSF) Grant AST 22-06053.

Our research made use of Parallelized rotation and flipping INvariant Kohonen-maps or \texttt{PINK}\footnote{\texttt{PINK}: \url{https://github.com/HITS-AIN/PINK}; \citealp{Polsterer}} as well as the python package \texttt{PYINK}\footnote{\texttt{PYINK}: \url{https://github.com/tjgalvin/pyink}}. We would like to thank Dr Tim Galvin for his help and advice on the installation and usage of \texttt{PINK}.

This work made use of Astropy:\footnote{http://www.astropy.org} a community-developed core Python package and an ecosystem of tools and resources for astronomy \citep{astropy:2013, astropy:2018, astropy:2022} as well as the Astropy affiliated package Astroquery (\citealp{astroquery}). Other packages used in this paper were Matplotlib (\citealp{Matplotlib}) and NumPy (\citealp{Numpy}).

This scientific work uses data obtained from Inyarrimanha Ilgari Bundara / the Murchison Radio-astronomy Observatory. We acknowledge the Wajarri Yamaji People as the Traditional Owners and native title holders of the Observatory site. CSIRO’s ASKAP radio telescope is part of the Australia Telescope National Facility (https://ror.org/05qajvd42). Operation of ASKAP is funded by the Australian Government with support from the National Collaborative Research Infrastructure Strategy. ASKAP uses the resources of the Pawsey Supercomputing Research Centre. Establishment of ASKAP, Inyarrimanha Ilgari Bundara, the CSIRO Murchison Radio-astronomy Observatory and the Pawsey Supercomputing Research Centre are initiatives of the Australian Government, with support from the Government of Western Australia and the Science and Industry Endowment Fund. This paper includes archived data obtained through the CSIRO ASKAP Science Data Archive, CASDA (https://data.csiro.au).

\end{acknowledgement}

%%%%%%%%%%%%%%%%%%%%%%%%%%%%%%%%%%%%%%%%%%%%%

% PASA uses footnotes, not endnotes. \endnote in this template will behave like \footnote; and \printendnotes will not output anything.
% \printendnotes

\bibliography{bib}

\begin{thebibliography}{}
\expandafter\ifx\csname natexlab\endcsname\relax\def\natexlab#1{#1}\fi

\bibitem[{{Abolfathi} {et~al.}(2018){Abolfathi}, {Aguado}, {Aguilar}, {Allende
  Prieto}, {Almeida}, {Ananna}, {Anders}, {Anderson}, {Andrews}, {Anguiano},
  {Arag{\'o}n-Salamanca}, {Argudo-Fern{\'a}ndez}, {Armengaud}, {Ata},
  {Aubourg}, {Avila-Reese}, {Badenes}, {Bailey}, {Balland}, {Barger},
  {Barrera-Ballesteros}, {Bartosz}, {Bastien}, {Bates}, {Baumgarten},
  {Bautista}, {Beaton}, {Beers}, {Belfiore}, {Bender}, {Bernardi}, {Bershady},
  {Beutler}, {Bird}, {Bizyaev}, {Blanc}, {Blanton}, {Blomqvist}, {Bolton},
  {Boquien}, {Borissova}, {Bovy}, {Bradna Diaz}, {Brandt}, {Brinkmann},
  {Brownstein}, {Bundy}, {Burgasser}, {Burtin}, {Busca}, {Ca{\~n}as},
  {Cano-D{\'\i}az}, {Cappellari}, {Carrera}, {Casey}, {Cervantes Sodi}, {Chen},
  {Cherinka}, {Chiappini}, {Choi}, {Chojnowski}, {Chuang}, {Chung}, {Clerc},
  {Cohen}, {Comerford}, {Comparat}, {Correa do Nascimento}, {da Costa},
  {Cousinou}, {Covey}, {Crane}, {Cruz-Gonzalez}, {Cunha}, {da Silva Ilha},
  {Damke}, {Darling}, {Davidson}, {Dawson}, {de Icaza Lizaola}, {de la
  Macorra}, {de la Torre}, {De Lee}, {de Sainte Agathe}, {Deconto Machado},
  {Dell'Agli}, {Delubac}, {Diamond-Stanic}, {Donor}, {Downes}, {Drory}, {du Mas
  des Bourboux}, {Duckworth}, {Dwelly}, {Dyer}, {Ebelke}, {Davis Eigenbrot},
  {Eisenstein}, {Elsworth}, {Emsellem}, {Eracleous}, {Erfanianfar},
  {Escoffier}, {Fan}, {Fern{\'a}ndez Alvar}, {Fernandez-Trincado}, {Fernando
  Cirolini}, {Feuillet}, {Finoguenov}, {Fleming}, {Font-Ribera}, {Freischlad},
  {Frinchaboy}, {Fu}, {G{\'o}mez Maqueo Chew}, {Galbany}, {Garc{\'\i}a
  P{\'e}rez}, {Garcia-Dias}, {Garc{\'\i}a-Hern{\'a}ndez}, {Garma Oehmichen},
  {Gaulme}, {Gelfand}, {Gil-Mar{\'\i}n}, {Gillespie}, {Goddard}, {Gonz{\'a}lez
  Hern{\'a}ndez}, {Gonzalez-Perez}, {Grabowski}, {Green}, {Grier}, {Gueguen},
  {Guo}, {Guy}, {Hagen}, {Hall}, {Harding}, {Hasselquist}, {Hawley}, {Hayes},
  {Hearty}, {Hekker}, {Hernandez}, {Hernandez Toledo}, {Hogg},
  {Holley-Bockelmann}, {Holtzman}, {Hou}, {Hsieh}, {Hunt}, {Hutchinson},
  {Hwang}, {Jimenez Angel}, {Johnson}, {Jones}, {J{\"o}nsson}, {Jullo}, {Khan},
  {Kinemuchi}, {Kirkby}, {Kirkpatrick}, {Kitaura}, {Knapp}, {Kneib},
  {Kollmeier}, {Lacerna}, {Lane}, {Lang}, {Law}, {Le Goff}, {Lee}, {Li}, {Li},
  {Lian}, {Liang}, {Lima}, {Lin}, {Long}, {Lucatello}, {Lundgren}, {Mackereth},
  {MacLeod}, {Mahadevan}, {Maia}, {Majewski}, {Manchado}, {Maraston},
  {Mariappan}, {Marques-Chaves}, {Masseron}, {Masters}, {McDermid}, {McGreer},
  {Melendez}, {Meneses-Goytia}, {Merloni}, {Merrifield}, {Meszaros}, {Meza},
  {Minchev}, {Minniti}, {Mueller}, {Muller-Sanchez}, {Muna}, {Mu{\~n}oz},
  {Myers}, {Nair}, {Nandra}, {Ness}, {Newman}, {Nichol}, {Nidever},
  {Nitschelm}, {Noterdaeme}, {O'Connell}, {Oelkers}, {Oravetz}, {Oravetz},
  {Ort{\'\i}z}, {Osorio}, {Pace}, {Padilla}, {Palanque-Delabrouille},
  {Palicio}, {Pan}, {Pan}, {Parikh}, {P{\^a}ris}, {Park}, {Peirani},
  {Pellejero-Ibanez}, {Penny}, {Percival}, {Perez-Fournon}, {Petitjean},
  {Pieri}, {Pinsonneault}, {Pisani}, {Prada}, {Prakash}, {Queiroz}, {Raddick},
  {Raichoor}, {Barboza Rembold}, {Richstein}, {Riffel}, {Riffel}, {Rix},
  {Robin}, {Rodr{\'\i}guez Torres}, {Rom{\'a}n-Z{\'u}{\~n}iga}, {Ross},
  {Rossi}, {Ruan}, {Ruggeri}, {Ruiz}, {Salvato}, {S{\'a}nchez}, {S{\'a}nchez},
  {Sanchez Almeida}, {S{\'a}nchez-Gallego}, {Santana Rojas}, {Santiago},
  {Schiavon}, {Schimoia}, {Schlafly}, {Schlegel}, {Schneider}, {Schuster},
  {Schwope}, {Seo}, {Serenelli}, {Shen}, {Shen}, {Shetrone}, {Shull}, {Silva
  Aguirre}, {Simon}, {Skrutskie}, {Slosar}, {Smethurst}, {Smith}, {Sobeck},
  {Somers}, {Souter}, {Souto}, {Spindler}, {Stark}, {Stassun}, {Steinmetz},
  {Stello}, {Storchi-Bergmann}, {Streblyanska}, {Stringfellow}, {Su{\'a}rez},
  {Sun}, {Szigeti}, {Taghizadeh-Popp}, {Talbot}, {Tang}, {Tao}, {Tayar},
  {Tembe}, {Teske}, {Thakar}, {Thomas}, {Tissera}, {Tojeiro}, {Tremonti},
  {Troup}, {Urry}, {Valenzuela}, {van den Bosch}, {Vargas-Gonz{\'a}lez},
  {Vargas-Maga{\~n}a}, {Vazquez}, {Villanova}, {Vogt}, {Wake}, {Wang},
  {Weaver}, {Weijmans}, {Weinberg}, {Westfall}, {Whelan}, {Wilcots}, {Wild},
  {Williams}, {Wilson}, {Wood-Vasey}, {Wylezalek}, {Xiao}, {Yan}, {Yang},
  {Ybarra}, {Y{\`e}che}, {Zakamska}, {Zamora}, {Zarrouk}, {Zasowski}, {Zhang},
  {Zhao}, {Zhao}, {Zheng}, {Zheng}, {Zhou}, {Zhu}, {Zinn}, \& {Zou}}]{SDSS14}
{Abolfathi}, B., {Aguado}, D.~S., {Aguilar}, G., {et~al.} 2018, \apjs, 235, 42

\bibitem[{Aniyan \& Thorat(2017)}]{Aniyan}
Aniyan, A.~K., \& Thorat, K. 2017, The Astrophysical Journal Supplement Series,
  230, 20

\bibitem[{{Astropy Collaboration} {et~al.}(2013){Astropy Collaboration},
  {Robitaille}, {Tollerud}, {Greenfield}, {Droettboom}, {Bray}, {Aldcroft},
  {Davis}, {Ginsburg}, {Price-Whelan}, {Kerzendorf}, {Conley}, {Crighton},
  {Barbary}, {Muna}, {Ferguson}, {Grollier}, {Parikh}, {Nair}, {Unther},
  {Deil}, {Woillez}, {Conseil}, {Kramer}, {Turner}, {Singer}, {Fox}, {Weaver},
  {Zabalza}, {Edwards}, {Azalee Bostroem}, {Burke}, {Casey}, {Crawford},
  {Dencheva}, {Ely}, {Jenness}, {Labrie}, {Lim}, {Pierfederici}, {Pontzen},
  {Ptak}, {Refsdal}, {Servillat}, \& {Streicher}}]{astropy:2013}
{Astropy Collaboration}, {Robitaille}, T.~P., {Tollerud}, E.~J., {et~al.} 2013,
  \aap, 558, A33

\bibitem[{{Astropy Collaboration} {et~al.}(2018){Astropy Collaboration},
  {Price-Whelan}, {Sip{\H{o}}cz}, {G{\"u}nther}, {Lim}, {Crawford}, {Conseil},
  {Shupe}, {Craig}, {Dencheva}, {Ginsburg}, {Vand erPlas}, {Bradley},
  {P{\'e}rez-Su{\'a}rez}, {de Val-Borro}, {Aldcroft}, {Cruz}, {Robitaille},
  {Tollerud}, {Ardelean}, {Babej}, {Bach}, {Bachetti}, {Bakanov}, {Bamford},
  {Barentsen}, {Barmby}, {Baumbach}, {Berry}, {Biscani}, {Boquien}, {Bostroem},
  {Bouma}, {Brammer}, {Bray}, {Breytenbach}, {Buddelmeijer}, {Burke},
  {Calderone}, {Cano Rodr{\'\i}guez}, {Cara}, {Cardoso}, {Cheedella}, {Copin},
  {Corrales}, {Crichton}, {D'Avella}, {Deil}, {Depagne}, {Dietrich}, {Donath},
  {Droettboom}, {Earl}, {Erben}, {Fabbro}, {Ferreira}, {Finethy}, {Fox},
  {Garrison}, {Gibbons}, {Goldstein}, {Gommers}, {Greco}, {Greenfield},
  {Groener}, {Grollier}, {Hagen}, {Hirst}, {Homeier}, {Horton}, {Hosseinzadeh},
  {Hu}, {Hunkeler}, {Ivezi{\'c}}, {Jain}, {Jenness}, {Kanarek}, {Kendrew},
  {Kern}, {Kerzendorf}, {Khvalko}, {King}, {Kirkby}, {Kulkarni}, {Kumar},
  {Lee}, {Lenz}, {Littlefair}, {Ma}, {Macleod}, {Mastropietro}, {McCully},
  {Montagnac}, {Morris}, {Mueller}, {Mumford}, {Muna}, {Murphy}, {Nelson},
  {Nguyen}, {Ninan}, {N{\"o}the}, {Ogaz}, {Oh}, {Parejko}, {Parley}, {Pascual},
  {Patil}, {Patil}, {Plunkett}, {Prochaska}, {Rastogi}, {Reddy Janga},
  {Sabater}, {Sakurikar}, {Seifert}, {Sherbert}, {Sherwood-Taylor}, {Shih},
  {Sick}, {Silbiger}, {Singanamalla}, {Singer}, {Sladen}, {Sooley},
  {Sornarajah}, {Streicher}, {Teuben}, {Thomas}, {Tremblay}, {Turner},
  {Terr{\'o}n}, {van Kerkwijk}, {de la Vega}, {Watkins}, {Weaver}, {Whitmore},
  {Woillez}, {Zabalza}, \& {Astropy Contributors}}]{astropy:2018}
{Astropy Collaboration}, {Price-Whelan}, A.~M., {Sip{\H{o}}cz}, B.~M., {et~al.}
  2018, \aj, 156, 123

\bibitem[{{Astropy Collaboration} {et~al.}(2022){Astropy Collaboration},
  {Price-Whelan}, {Lim}, {Earl}, {Starkman}, {Bradley}, {Shupe}, {Patil},
  {Corrales}, {Brasseur}, {N{"o}the}, {Donath}, {Tollerud}, {Morris},
  {Ginsburg}, {Vaher}, {Weaver}, {Tocknell}, {Jamieson}, {van Kerkwijk},
  {Robitaille}, {Merry}, {Bachetti}, {G{"u}nther}, {Aldcroft},
  {Alvarado-Montes}, {Archibald}, {B{'o}di}, {Bapat}, {Barentsen}, {Baz{'a}n},
  {Biswas}, {Boquien}, {Burke}, {Cara}, {Cara}, {Conroy}, {Conseil}, {Craig},
  {Cross}, {Cruz}, {D'Eugenio}, {Dencheva}, {Devillepoix}, {Dietrich},
  {Eigenbrot}, {Erben}, {Ferreira}, {Foreman-Mackey}, {Fox}, {Freij}, {Garg},
  {Geda}, {Glattly}, {Gondhalekar}, {Gordon}, {Grant}, {Greenfield}, {Groener},
  {Guest}, {Gurovich}, {Handberg}, {Hart}, {Hatfield-Dodds}, {Homeier},
  {Hosseinzadeh}, {Jenness}, {Jones}, {Joseph}, {Kalmbach}, {Karamehmetoglu},
  {Ka{l}uszy{'n}ski}, {Kelley}, {Kern}, {Kerzendorf}, {Koch}, {Kulumani},
  {Lee}, {Ly}, {Ma}, {MacBride}, {Maljaars}, {Muna}, {Murphy}, {Norman},
  {O'Steen}, {Oman}, {Pacifici}, {Pascual}, {Pascual-Granado}, {Patil},
  {Perren}, {Pickering}, {Rastogi}, {Roulston}, {Ryan}, {Rykoff}, {Sabater},
  {Sakurikar}, {Salgado}, {Sanghi}, {Saunders}, {Savchenko}, {Schwardt},
  {Seifert-Eckert}, {Shih}, {Jain}, {Shukla}, {Sick}, {Simpson},
  {Singanamalla}, {Singer}, {Singhal}, {Sinha}, {Sip{H{o}}cz}, {Spitler},
  {Stansby}, {Streicher}, {{{S}}umak}, {Swinbank}, {Taranu}, {Tewary},
  {Tremblay}, {Val-Borro}, {Van Kooten}, {Vasovi{'c}}, {Verma}, {de Miranda
  Cardoso}, {Williams}, {Wilson}, {Winkel}, {Wood-Vasey}, {Xue}, {Yoachim},
  {Zhang}, {Zonca}, \& {Astropy Project Contributors}}]{astropy:2022}
{Astropy Collaboration}, {Price-Whelan}, A.~M., {Lim}, P.~L., {et~al.} 2022,
  \apj, 935, 167

\bibitem[{Banfield {et~al.}(2015)Banfield, Wong, Willett, Norris, Rudnick,
  Shabala, Simmons, Snyder, Garon, Seymour, \& et~al.}]{Banfield}
Banfield, J.~K., Wong, O.~I., Willett, K.~W., {et~al.} 2015, Monthly Notices of
  the Royal Astronomical Society, 453, 2327–2341

\bibitem[{Baron(2019)}]{Baron}
Baron, D. 2019, Machine Learning in Astronomy: a practical overview,
  doi:\url{10.48550/ARXIV.1904.07248}

\bibitem[{{Becker} {et~al.}(1995){Becker}, {White}, \& {Helfand}}]{FIRST}
{Becker}, R.~H., {White}, R.~L., \& {Helfand}, D.~J. 1995, \apj, 450, 559

\bibitem[{Bock {et~al.}(1999)Bock, Large, \& Sadler}]{SUMSS}
Bock, D. C.-J., Large, M.~I., \& Sadler, E.~M. 1999, The Astronomical Journal,
  117, 1578

\bibitem[{{Bowles} {et~al.}(2023){Bowles}, {Tang}, {Vardoulaki}, {Alexander},
  {Luo}, {Rudnick}, {Walmsley}, {Porter}, {Scaife}, {Slijepcevic}, {Adams},
  {Drabent}, {Dugdale}, {G{\"u}rkan}, {Hopkins}, {Jimenez-Andrade}, {Leahy},
  {Norris}, {Rahman}, {Ouyang}, {Segal}, {Shabala}, \& {Wong}}]{Bowles2023}
{Bowles}, M., {Tang}, H., {Vardoulaki}, E., {et~al.} 2023, \mnras, 522, 2584

\bibitem[{{Condon}(1992)}]{Condon}
{Condon}, J.~J. 1992, \araa, 30, 575

\bibitem[{{Condon} {et~al.}(1998){Condon}, {Cotton}, {Greisen}, {Yin},
  {Perley}, {Taylor}, \& {Broderick}}]{NVSS}
{Condon}, J.~J., {Cotton}, W.~D., {Greisen}, E.~W., {et~al.} 1998, \aj, 115,
  1693

\bibitem[{{Edge} {et~al.}(1959){Edge}, {Shakeshaft}, {McAdam}, {Baldwin}, \&
  {Archer}}]{3C}
{Edge}, D.~O., {Shakeshaft}, J.~R., {McAdam}, W.~B., {Baldwin}, J.~E., \&
  {Archer}, S. 1959, \memras, 68, 37

\bibitem[{{Ekers}(1969)}]{Parkes}
{Ekers}, J.~A. 1969, Australian Journal of Physics Astrophysical Supplement, 7,
  3

\bibitem[{Everitt {et~al.}(2011)Everitt, Landau, Leese, \& Stahl}]{GMM}
Everitt, B., Landau, S., Leese, M., \& Stahl, D. 2011, Cluster analysis, 5th
  edn. (Wiley)

\bibitem[{{Fanaroff} \& {Riley}(1974)}]{Fanaroff&Riley}
{Fanaroff}, B.~L., \& {Riley}, J.~M. 1974, \mnras, 167, 31P

\bibitem[{Galvin {et~al.}(2019)Galvin, Huynh, Norris, Wang, Hopkins, Wong,
  Shabala, Rudnick, Alger, \& Polsterer}]{Galvin2019}
Galvin, T.~J., Huynh, M., Norris, R.~P., {et~al.} 2019, Publications of the
  Astronomical Society of the Pacific, 131, 108009

\bibitem[{Galvin {et~al.}(2020)Galvin, Huynh, Norris, Wang, Hopkins, Polsterer,
  Ralph, O'Brien, \& Heald}]{Galvin2020}
Galvin, T.~J., Huynh, M.~T., Norris, R.~P., {et~al.} 2020, Monthly Notices of
  the Royal Astronomical Society, 497, 2730

\bibitem[{{Ginsburg} {et~al.}(2019){Ginsburg}, {Sip{\H{o}}cz}, {Brasseur},
  {Cowperthwaite}, {Craig}, {Deil}, {Guillochon}, {Guzman}, {Liedtke}, {Lian
  Lim}, {Lockhart}, {Mommert}, {Morris}, {Norman}, {Parikh}, {Persson},
  {Robitaille}, {Segovia}, {Singer}, {Tollerud}, {de Val-Borro}, {Valtchanov},
  {Woillez}, {Astroquery Collaboration}, \& {a subset of astropy
  Collaboration}}]{astroquery}
{Ginsburg}, A., {Sip{\H{o}}cz}, B.~M., {Brasseur}, C.~E., {et~al.} 2019, \aj,
  157, 98

\bibitem[{{Gordon} {et~al.}(2023){Gordon}, {Rudnick}, {Andernach}, {Morabito},
  {O'Dea}, {Achong}, {Baum}, {Bayona-Figueroa}, {Hooper}, {Mingo}, {Morris}, \&
  {Vantyghem}}]{Gordon2023}
{Gordon}, Y.~A., {Rudnick}, L., {Andernach}, H., {et~al.} 2023, \apjs, 267, 37

\bibitem[{{G{\"u}rkan} {et~al.}(2022){G{\"u}rkan}, {Prandoni}, {O'Brien},
  {Raja}, {Marchetti}, {Vaccari}, {Driver}, {Taylor}, {Franzen}, {Brown},
  {Shabala}, {Andernach}, {Hopkins}, {Norris}, {Leahy}, {Bilicki},
  {Farajollahi}, {Galvin}, {Heald}, {Koribalski}, {An}, \&
  {Warhurst}}]{Gurkan2022}
{G{\"u}rkan}, G., {Prandoni}, I., {O'Brien}, A., {et~al.} 2022, \mnras, 512,
  6104

\bibitem[{Hale {et~al.}(2021)Hale, McConnell, Thomson, Lenc, Heald, Hotan,
  Leung, Moss, Murphy, Pritchard, Sadler, Stewart, \& Whiting}]{Hale2021}
Hale, C.~L., McConnell, D., Thomson, A. J.~M., {et~al.} 2021, Publications of
  the Astronomical Society of Australia, 38, doi:\url{10.1017/pasa.2021.47}

\bibitem[{{Harris} {et~al.}(2020){Harris}, {Millman}, {van der Walt},
  {Gommers}, {Virtanen}, {Cournapeau}, {Wieser}, {Taylor}, {Berg}, {Smith},
  {Kern}, {Picus}, {Hoyer}, {van Kerkwijk}, {Brett}, {Haldane}, {del R{\'\i}o},
  {Wiebe}, {Peterson}, {G{\'e}rard-Marchant}, {Sheppard}, {Reddy}, {Weckesser},
  {Abbasi}, {Gohlke}, \& {Oliphant}}]{Numpy}
{Harris}, C.~R., {Millman}, K.~J., {van der Walt}, S.~J., {et~al.} 2020, \nat,
  585, 357

\bibitem[{{Hotan} {et~al.}(2021){Hotan}, {Bunton}, {Chippendale}, {Whiting},
  {Tuthill}, {Moss}, {McConnell}, {Amy}, {Huynh}, {Allison}, {Anderson},
  {Bannister}, {Bastholm}, {Beresford}, {Bock}, {Bolton}, {Chapman}, {Chow},
  {Collier}, {Cooray}, {Cornwell}, {Diamond}, {Edwards}, {Feain}, {Franzen},
  {George}, {Gupta}, {Hampson}, {Harvey-Smith}, {Hayman}, {Heywood}, {Jacka},
  {Jackson}, {Jackson}, {Jeganathan}, {Johnston}, {Kesteven}, {Kleiner},
  {Koribalski}, {Lee-Waddell}, {Lenc}, {Lensson}, {Mackay}, {Mahony},
  {McClure-Griffiths}, {McConigley}, {Mirtschin}, {Ng}, {Norris}, {Pearce},
  {Phillips}, {Pilawa}, {Raja}, {Reynolds}, {Roberts}, {Roxby}, {Sadler},
  {Shields}, {Schinckel}, {Serra}, {Shaw}, {Sweetnam}, {Troup}, {Tzioumis},
  {Voronkov}, \& {Westmeier}}]{Hotan}
{Hotan}, A.~W., {Bunton}, J.~D., {Chippendale}, A.~P., {et~al.} 2021, \pasa,
  38, e009

\bibitem[{Hunter(2007)}]{Matplotlib}
Hunter, J.~D. 2007, Computing in Science \& Engineering, 9, 90

\bibitem[{{Hurley-Walker} {et~al.}(2017){Hurley-Walker}, {Callingham},
  {Hancock}, {Franzen}, {Hindson}, {Kapi{\'n}ska}, {Morgan}, {Offringa},
  {Wayth}, {Wu}, {Zheng}, {Murphy}, {Bell}, {Dwarakanath}, {For}, {Gaensler},
  {Johnston-Hollitt}, {Lenc}, {Procopio}, {Staveley-Smith}, {Ekers}, {Bowman},
  {Briggs}, {Cappallo}, {Deshpande}, {Greenhill}, {Hazelton}, {Kaplan},
  {Lonsdale}, {McWhirter}, {Mitchell}, {Morales}, {Morgan}, {Oberoi}, {Ord},
  {Prabu}, {Shankar}, {Srivani}, {Subrahmanyan}, {Tingay}, {Webster},
  {Williams}, \& {Williams}}]{GLEAM}
{Hurley-Walker}, N., {Callingham}, J.~R., {Hancock}, P.~J., {et~al.} 2017,
  \mnras, 464, 1146

\bibitem[{Ikotun {et~al.}(2022)Ikotun, Ezugwu, Abualigah, Abuhaija, \&
  Heming}]{kmeans}
Ikotun, A., Ezugwu, A., Abualigah, L., Abuhaija, B., \& Heming, J. 2022,
  Information Sciences, 622, doi:\url{10.1016/j.ins.2022.11.139}

\bibitem[{Intema {et~al.}(2017)Intema, Jagannathan, Mooley, \& Frail}]{TGSS}
Intema, H.~T., Jagannathan, P., Mooley, K.~P., \& Frail, D.~A. 2017, Astronomy
  \& Astrophysics, 598, A78

\bibitem[{Johnston {et~al.}(2008)Johnston, Taylor, Bailes, Bartel, Baugh,
  Bietenholz, Blake, Braun, Brown, Chatterjee, Darling, Deller, Dodson,
  Edwards, Ekers, Ellingsen, Feain, Gaensler, Haverkorn, Hobbs, Hopkins,
  Jackson, James, Joncas, Kaspi, Kilborn, Koribalski, Kothes, Landecker, Lenc,
  Lovell, Macquart, Manchester, Matthews, McClure-Griffiths, Norris, Pen,
  Phillips, Power, Protheroe, Sadler, Schmidt, Stairs, Staveley-Smith, Stil,
  Tingay, Tzioumis, Walker, Wall, \& Wolleben}]{Johnston2008}
Johnston, S., Taylor, R., Bailes, M., {et~al.} 2008, Experimental Astronomy,
  22, 151

\bibitem[{Jolliffe \& Cadima(2016)}]{PCA}
Jolliffe, I., \& Cadima, J. 2016, Philosophical Transactions of the Royal
  Society A: Mathematical, Physical and Engineering Sciences, 374, 20150202

\bibitem[{{Kellermann} {et~al.}(1989){Kellermann}, {Sramek}, {Schmidt},
  {Shaffer}, \& {Green}}]{Kellermann1989}
{Kellermann}, K.~I., {Sramek}, R., {Schmidt}, M., {Shaffer}, D.~B., \& {Green},
  R. 1989, \aj, 98, 1195

\bibitem[{Kohonen(1990)}]{Kohonen1990}
Kohonen, T. 1990, Proceedings of the IEEE, 78, 1464

\bibitem[{{Kohonen}(2001)}]{Kohonen}
{Kohonen}, T. 2001, {Self-Organizing Maps}, 3rd edn. (Springer)

\bibitem[{{Kormendy} \& {Ho}(2013)}]{Kormendy}
{Kormendy}, J., \& {Ho}, L.~C. 2013, \araa, 51, 511

\bibitem[{{Lacy} {et~al.}(2020){Lacy}, {Baum}, {Chandler}, {Chatterjee},
  {Clarke}, {Deustua}, {English}, {Farnes}, {Gaensler}, {Gugliucci},
  {Hallinan}, {Kent}, {Kimball}, {Law}, {Lazio}, {Marvil}, {Mao}, {Medlin},
  {Mooley}, {Murphy}, {Myers}, {Osten}, {Richards}, {Rosolowsky}, {Rudnick},
  {Schinzel}, {Sivakoff}, {Sjouwerman}, {Taylor}, {White}, {Wrobel},
  {Andernach}, {Beasley}, {Berger}, {Bhatnager}, {Birkinshaw}, {Bower},
  {Brandt}, {Brown}, {Burke-Spolaor}, {Butler}, {Comerford}, {Demorest}, {Fu},
  {Giacintucci}, {Golap}, {G{\"u}th}, {Hales}, {Hiriart}, {Hodge}, {Horesh},
  {Ivezi{\'c}}, {Jarvis}, {Kamble}, {Kassim}, {Liu}, {Loinard}, {Lyons},
  {Masters}, {Mezcua}, {Moellenbrock}, {Mroczkowski}, {Nyland}, {O'Dea},
  {O'Sullivan}, {Peters}, {Radford}, {Rao}, {Robnett}, {Salcido}, {Shen},
  {Sobotka}, {Witz}, {Vaccari}, {van Weeren}, {Vargas}, {Williams}, \&
  {Yoon}}]{VLASS}
{Lacy}, M., {Baum}, S.~A., {Chandler}, C.~J., {et~al.} 2020, \pasp, 132, 035001

\bibitem[{{Lara} {et~al.}(2001){Lara}, {Cotton}, {Feretti}, {Giovannini},
  {Marcaide}, {M{\'a}rquez}, \& {Venturi}}]{Lara2001}
{Lara}, L., {Cotton}, W.~D., {Feretti}, L., {et~al.} 2001, \aap, 370, 409

\bibitem[{{Lukic} {et~al.}(2018){Lukic}, {Br{\"u}ggen}, {Banfield}, {Wong},
  {Rudnick}, {Norris}, \& {Simmons}}]{Lukic2018}
{Lukic}, V., {Br{\"u}ggen}, M., {Banfield}, J.~K., {et~al.} 2018, \mnras, 476,
  246

\bibitem[{McConnell {et~al.}(2020)McConnell, Hale, Lenc, Banfield, Heald,
  Hotan, Leung, Moss, Murphy, O'Brien, Pritchard, Raja, Sadler, Stewart,
  Thomson, Whiting, Allison, Amy, Anderson, Ball, Bannister, Bell, Bock,
  Bolton, Bunton, Chippendale, Collier, Cooray, Cornwell, Diamond, Edwards,
  Gupta, Hayman, Heywood, Jackson, Koribalski, Lee-Waddell, McClure-Griffiths,
  Ng, Norris, Phillips, Reynolds, Roxby, Schinckel, Shields, Tremblay,
  Tzioumis, Voronkov, \& Westmeier}]{McConnell2020}
McConnell, D., Hale, C.~L., Lenc, E., {et~al.} 2020, Publications of the
  Astronomical Society of Australia, 37, doi:\url{10.1017/pasa.2020.41}

\bibitem[{{Mohan} \& {Rafferty}(2015)}]{PyBDSF}
{Mohan}, N., \& {Rafferty}, D. 2015, {PyBDSF: Python Blob Detection and Source
  Finder}, Astrophysics Source Code Library, record ascl:1502.007

\bibitem[{{Mostert} {et~al.}(2021){Mostert}, {Duncan}, {R{\"o}ttgering},
  {Polsterer}, {Best}, {Brienza}, {Br{\"u}ggen}, {Hardcastle}, {Jurlin},
  {Mingo}, {Morganti}, {Shimwell}, {Smith}, \& {Williams}}]{Mostert2021}
{Mostert}, R. I.~J., {Duncan}, K.~J., {R{\"o}ttgering}, H. J.~A., {et~al.}
  2021, \aap, 645, A89

\bibitem[{Norris(2017)}]{Norris2017}
Norris, R.~P. 2017, Nature Astronomy, 1, 671–678

\bibitem[{{Norris} {et~al.}(2011){Norris}, {Hopkins}, {Afonso}, {Brown},
  {Condon}, {Dunne}, {Feain}, {Hollow}, {Jarvis}, {Johnston-Hollitt}, {Lenc},
  {Middelberg}, {Padovani}, {Prandoni}, {Rudnick}, {Seymour}, {Umana},
  {Andernach}, {Alexander}, {Appleton}, {Bacon}, {Banfield}, {Becker}, {Brown},
  {Ciliegi}, {Jackson}, {Eales}, {Edge}, {Gaensler}, {Giovannini}, {Hales},
  {Hancock}, {Huynh}, {Ibar}, {Ivison}, {Kennicutt}, {Kimball}, {Koekemoer},
  {Koribalski}, {L{\'o}pez-S{\'a}nchez}, {Mao}, {Murphy}, {Messias},
  {Pimbblet}, {Raccanelli}, {Randall}, {Reiprich}, {Roseboom},
  {R{\"o}ttgering}, {Saikia}, {Sharp}, {Slee}, {Smail}, {Thompson}, {Urquhart},
  {Wall}, \& {Zhao}}]{EMU2}
{Norris}, R.~P., {Hopkins}, A.~M., {Afonso}, J., {et~al.} 2011, \pasa, 28, 215

\bibitem[{{Norris} {et~al.}(2021){Norris}, {Marvil}, {Collier}, {Kapi{\'n}ska},
  {O'Brien}, {Rudnick}, {Andernach}, {Asorey}, {Brown}, {Br{\"u}ggen},
  {Crawford}, {English}, {Rahman}, {Filipovi{\'c}}, {Gordon}, {G{\"u}rkan},
  {Hale}, {Hopkins}, {Huynh}, {HyeongHan}, {James Jee}, {Koribalski}, {Lenc},
  {Luken}, {Parkinson}, {Prandoni}, {Raja}, {Reiprich}, {Riseley}, {Shabala},
  {Sheil}, {Vernstrom}, {Whiting}, {Allison}, {Anderson}, {Ball}, {Bell},
  {Bunton}, {Galvin}, {Gupta}, {Hotan}, {Jacka}, {Macgregor}, {Mahony}, {Maio},
  {Moss}, {Pandey-Pommier}, \& {Voronkov}}]{Emupilot}
{Norris}, R.~P., {Marvil}, J., {Collier}, J.~D., {et~al.} 2021, \pasa, 38, e046

\bibitem[{{Ochsenbein} {et~al.}(2000){Ochsenbein}, {Bauer}, \&
  {Marcout}}]{VizieR}
{Ochsenbein}, F., {Bauer}, P., \& {Marcout}, J. 2000, \aaps, 143, 23

\bibitem[{Padovani {et~al.}(2017)Padovani, Alexander, Assef, Marco, Giommi,
  Hickox, Richards, Smol{\v{c}}i{\'{c}}, Hatziminaoglou, Mainieri, \&
  Salvato}]{Padovani2017}
Padovani, P., Alexander, D.~M., Assef, R.~J., {et~al.} 2017, The Astronomy and
  Astrophysics Review, 25, doi:\url{10.1007/s00159-017-0102-9}

\bibitem[{{Perley} {et~al.}(2011){Perley}, {Chandler}, {Butler}, \&
  {Wrobel}}]{VLA}
{Perley}, R.~A., {Chandler}, C.~J., {Butler}, B.~J., \& {Wrobel}, J.~M. 2011,
  \apjl, 739, L1

\bibitem[{Polsterer {et~al.}(2016)Polsterer, Gieseke, Igel, Doser, \&
  Gianniotis}]{Polsterer}
Polsterer, K., Gieseke, F., Igel, C., Doser, B., \& Gianniotis, N. 2016, in
  24$^{th}$ European Symposium on Artificial Neural Networks, ESANN 2016,
  Bruges, Belgium, April 27-29, 2016, Bruges, Belgium, publication status:
  Published

\bibitem[{{Polsterer} {et~al.}(2015){Polsterer}, {Gieseke}, \&
  {Igel}}]{Polsterer2015}
{Polsterer}, K.~L., {Gieseke}, F., \& {Igel}, C. 2015, in Astronomical Society
  of the Pacific Conference Series, Vol. 495, Astronomical Data Analysis
  Software an Systems XXIV (ADASS XXIV), ed. A.~R. {Taylor} \& E.~{Rosolowsky},
  81

\bibitem[{{Proctor}(2016)}]{Proctor2016}
{Proctor}, D.~D. 2016, \apjs, 224, 18

\bibitem[{{Rengelink} {et~al.}(1997){Rengelink}, {Tang}, {de Bruyn}, {Miley},
  {Bremer}, {Roettgering}, \& {Bremer}}]{WENSS}
{Rengelink}, R.~B., {Tang}, Y., {de Bruyn}, A.~G., {et~al.} 1997, \aaps, 124,
  259

\bibitem[{{Rudnick}(2021)}]{Rudnick2021}
{Rudnick}, L. 2021, Galaxies, 9, 85

\bibitem[{{Sadler} {et~al.}(1989){Sadler}, {Jenkins}, \& {Kotanyi}}]{Sadler}
{Sadler}, E.~M., {Jenkins}, C.~R., \& {Kotanyi}, C.~G. 1989, \mnras, 240, 591

\bibitem[{{Savage} \& {Wall}(1976)}]{SavageWall}
{Savage}, A., \& {Wall}, J.~V. 1976, Australian Journal of Physics
  Astrophysical Supplement, 39, 39

\bibitem[{{Shimwell} {et~al.}(2017){Shimwell}, {R{\"o}ttgering}, {Best},
  {Williams}, {Dijkema}, {de Gasperin}, {Hardcastle}, {Heald}, {Hoang},
  {Horneffer}, {Intema}, {Mahony}, {Mandal}, {Mechev}, {Morabito}, {Oonk},
  {Rafferty}, {Retana-Montenegro}, {Sabater}, {Tasse}, {van Weeren},
  {Br{\"u}ggen}, {Brunetti}, {Chy{\.z}y}, {Conway}, {Haverkorn}, {Jackson},
  {Jarvis}, {McKean}, {Miley}, {Morganti}, {White}, {Wise}, {van Bemmel},
  {Beck}, {Brienza}, {Bonafede}, {Calistro Rivera}, {Cassano}, {Clarke},
  {Cseh}, {Deller}, {Drabent}, {van Driel}, {Engels}, {Falcke}, {Ferrari},
  {Fr{\"o}hlich}, {Garrett}, {Harwood}, {Heesen}, {Hoeft}, {Horellou},
  {Israel}, {Kapi{\'n}ska}, {Kunert-Bajraszewska}, {McKay}, {Mohan},
  {Orr{\'u}}, {Pizzo}, {Prandoni}, {Schwarz}, {Shulevski}, {Sipior}, {Smith},
  {Sridhar}, {Steinmetz}, {Stroe}, {Varenius}, {van der Werf}, {Zensus}, \&
  {Zwart}}]{LoTSS}
{Shimwell}, T.~W., {R{\"o}ttgering}, H.~J.~A., {Best}, P.~N., {et~al.} 2017,
  \aap, 598, A104

\bibitem[{{Sutherland} \& {Saunders}(1992)}]{Sutherland}
{Sutherland}, W., \& {Saunders}, W. 1992, \mnras, 259, 413

\bibitem[{Tingay {et~al.}(2013)Tingay, Goeke, Bowman, Emrich, Ord, Mitchell,
  Morales, Booler, Crosse, Wayth, Lonsdale, Tremblay, Pallot, Colegate,
  Wicenec, Kudryavtseva, Arcus, Barnes, Bernardi, Briggs, Burns, Bunton,
  Cappallo, Corey, Deshpande, Desouza, Gaensler, Greenhill, Hall, Hazelton,
  Herne, Hewitt, Johnston-Hollitt, Kaplan, Kasper, Kincaid, Koenig,
  Kratzenberg, Lynch, Mckinley, Mcwhirter, Morgan, Oberoi, Pathikulangara,
  Prabu, Remillard, Rogers, Roshi, Salah, Sault, Udaya-Shankar, Schlagenhaufer,
  Srivani, Stevens, Subrahmanyan, Waterson, Webster, Whitney, Williams,
  Williams, \& Wyithe}]{Tingay2013}
Tingay, S.~J., Goeke, R., Bowman, J.~D., {et~al.} 2013, Publications of the
  Astronomical Society of Australia, 30, doi:\url{10.1017/pasa.2012.007}

\bibitem[{Urry \& Padovani(1995)}]{Urry1995}
Urry, C.~M., \& Padovani, P. 1995, Publications of the Astronomical Society of
  the Pacific, 107, 803

\bibitem[{{van Haarlem} {et~al.}(2013){van Haarlem}, {Wise}, {Gunst}, {Heald},
  {McKean}, {Hessels}, {de Bruyn}, {Nijboer}, {Swinbank}, {Fallows},
  {Brentjens}, {Nelles}, {Beck}, {Falcke}, {Fender}, {H{\"o}randel},
  {Koopmans}, {Mann}, {Miley}, {R{\"o}ttgering}, {Stappers}, {Wijers},
  {Zaroubi}, {van den Akker}, {Alexov}, {Anderson}, {Anderson}, {van Ardenne},
  {Arts}, {Asgekar}, {Avruch}, {Batejat}, {B{\"a}hren}, {Bell}, {Bell}, {van
  Bemmel}, {Bennema}, {Bentum}, {Bernardi}, {Best}, {B{\^\i}rzan}, {Bonafede},
  {Boonstra}, {Braun}, {Bregman}, {Breitling}, {van de Brink}, {Broderick},
  {Broekema}, {Brouw}, {Br{\"u}ggen}, {Butcher}, {van Cappellen}, {Ciardi},
  {Coenen}, {Conway}, {Coolen}, {Corstanje}, {Damstra}, {Davies}, {Deller},
  {Dettmar}, {van Diepen}, {Dijkstra}, {Donker}, {Doorduin}, {Dromer}, {Drost},
  {van Duin}, {Eisl{\"o}ffel}, {van Enst}, {Ferrari}, {Frieswijk}, {Gankema},
  {Garrett}, {de Gasperin}, {Gerbers}, {de Geus}, {Grie{\ss}meier}, {Grit},
  {Gruppen}, {Hamaker}, {Hassall}, {Hoeft}, {Holties}, {Horneffer}, {van der
  Horst}, {van Houwelingen}, {Huijgen}, {Iacobelli}, {Intema}, {Jackson},
  {Jelic}, {de Jong}, {Juette}, {Kant}, {Karastergiou}, {Koers}, {Kollen},
  {Kondratiev}, {Kooistra}, {Koopman}, {Koster}, {Kuniyoshi}, {Kramer},
  {Kuper}, {Lambropoulos}, {Law}, {van Leeuwen}, {Lemaitre}, {Loose}, {Maat},
  {Macario}, {Markoff}, {Masters}, {McFadden}, {McKay-Bukowski}, {Meijering},
  {Meulman}, {Mevius}, {Middelberg}, {Millenaar}, {Miller-Jones}, {Mohan},
  {Mol}, {Morawietz}, {Morganti}, {Mulcahy}, {Mulder}, {Munk}, {Nieuwenhuis},
  {van Nieuwpoort}, {Noordam}, {Norden}, {Noutsos}, {Offringa}, {Olofsson},
  {Omar}, {Orr{\'u}}, {Overeem}, {Paas}, {Pandey-Pommier}, {Pandey}, {Pizzo},
  {Polatidis}, {Rafferty}, {Rawlings}, {Reich}, {de Reijer}, {Reitsma},
  {Renting}, {Riemers}, {Rol}, {Romein}, {Roosjen}, {Ruiter}, {Scaife}, {van
  der Schaaf}, {Scheers}, {Schellart}, {Schoenmakers}, {Schoonderbeek},
  {Serylak}, {Shulevski}, {Sluman}, {Smirnov}, {Sobey}, {Spreeuw}, {Steinmetz},
  {Sterks}, {Stiepel}, {Stuurwold}, {Tagger}, {Tang}, {Tasse}, {Thomas},
  {Thoudam}, {Toribio}, {van der Tol}, {Usov}, {van Veelen}, {van der Veen},
  {ter Veen}, {Verbiest}, {Vermeulen}, {Vermaas}, {Vocks}, {Vogt}, {de Vos},
  {van der Wal}, {van Weeren}, {Weggemans}, {Weltevrede}, {White}, {Wijnholds},
  {Wilhelmsson}, {Wucknitz}, {Yatawatta}, {Zarka}, {Zensus}, \& {van
  Zwieten}}]{vanHaarlem}
{van Haarlem}, M.~P., {Wise}, M.~W., {Gunst}, A.~W., {et~al.} 2013, \aap, 556,
  A2

\bibitem[{{Vantyghem} {et~al.}(2024){Vantyghem}, {Galvin}, {Sebastian},
  {O'Dea}, {Gordon}, {Boyce}, {Rudnick}, {Polsterer}, {Andernach},
  {Dionyssiou}, {Venkataraman}, {Norris}, {Baum}, {Wang}, \&
  {Huynh}}]{Vantyghem}
{Vantyghem}, A.~N., {Galvin}, T.~J., {Sebastian}, B., {et~al.} 2024, Astronomy
  and Computing, 47, 100824

\bibitem[{Wayth {et~al.}(2018)Wayth, Tingay, Trott, Emrich, Johnston-Hollitt,
  McKinley, Gaensler, Beardsley, Booler, Crosse, Franzen, Horsley, Kaplan,
  Kenney, Morales, Pallot, Sleap, Steele, Walker, Williams, Wu, Cairns,
  Filipovic, Johnston, Murphy, Quinn, Staveley-Smith, Webster, \&
  Wyithe}]{Wayth2018}
Wayth, R.~B., Tingay, S.~J., Trott, C.~M., {et~al.} 2018, Publications of the
  Astronomical Society of Australia, 35, doi:\url{10.1017/pasa.2018.37}

\bibitem[{{Williams} {et~al.}(2019){Williams}, {Hardcastle}, {Best}, {Sabater},
  {Croston}, {Duncan}, {Shimwell}, {R{\"o}ttgering}, {Nisbet}, {G{\"u}rkan},
  {Alegre}, {Cochrane}, {Goyal}, {Hale}, {Jackson}, {Jamrozy}, {Kondapally},
  {Kunert-Bajraszewska}, {Mahatma}, {Mingo}, {Morabito}, {Prandoni},
  {Roskowinski}, {Shulevski}, {Smith}, {Tasse}, {Urquhart}, {Webster}, {White},
  {Beswick}, {Callingham}, {Chy{\.z}y}, {de Gasperin}, {Harwood}, {Hoeft},
  {Iacobelli}, {McKean}, {Mechev}, {Miley}, {Schwarz}, \& {van
  Weeren}}]{Williams2019}
{Williams}, W.~L., {Hardcastle}, M.~J., {Best}, P.~N., {et~al.} 2019, \aap,
  622, A2

\bibitem[{{Windhorst} {et~al.}(1984){Windhorst}, {Kron}, \&
  {Koo}}]{Windhorst1984}
{Windhorst}, R.~A., {Kron}, R.~G., \& {Koo}, D.~C. 1984, \aaps, 58, 39

\bibitem[{{Windhorst} {et~al.}(1985){Windhorst}, {Miley}, {Owen}, {Kron}, \&
  {Koo}}]{Windhorst1985}
{Windhorst}, R.~A., {Miley}, G.~K., {Owen}, F.~N., {Kron}, R.~G., \& {Koo},
  D.~C. 1985, \apj, 289, 494

\bibitem[{{Wright} {et~al.}(2010){Wright}, {Eisenhardt}, {Mainzer}, {Ressler},
  {Cutri}, {Jarrett}, {Kirkpatrick}, {Padgett}, {McMillan}, {Skrutskie},
  {Stanford}, {Cohen}, {Walker}, {Mather}, {Leisawitz}, {Gautier}, {McLean},
  {Benford}, {Lonsdale}, {Blain}, {Mendez}, {Irace}, {Duval}, {Liu}, {Royer},
  {Heinrichsen}, {Howard}, {Shannon}, {Kendall}, {Walsh}, {Larsen}, {Cardon},
  {Schick}, {Schwalm}, {Abid}, {Fabinsky}, {Naes}, \& {Tsai}}]{Wright}
{Wright}, E.~L., {Eisenhardt}, P. R.~M., {Mainzer}, A.~K., {et~al.} 2010, \aj,
  140, 1868

\bibitem[{York {et~al.}(2000)York, Adelman, Anderson, Anderson, Annis, Bahcall,
  Bakken, Barkhouser, Bastian, Berman, \& et~al.}]{York2000}
York, D.~G., Adelman, J., Anderson, Jr., J.~E., {et~al.} 2000, The Astronomical
  Journal, 120, 1579–1587

\end{thebibliography}

\appendix

\section{BMU Euclidean Distances of individual neurons in SOM grid}
\label{appendix1}

\begin{figure*}%[!b]
    \centering
        \includegraphics[width=1.0\linewidth]{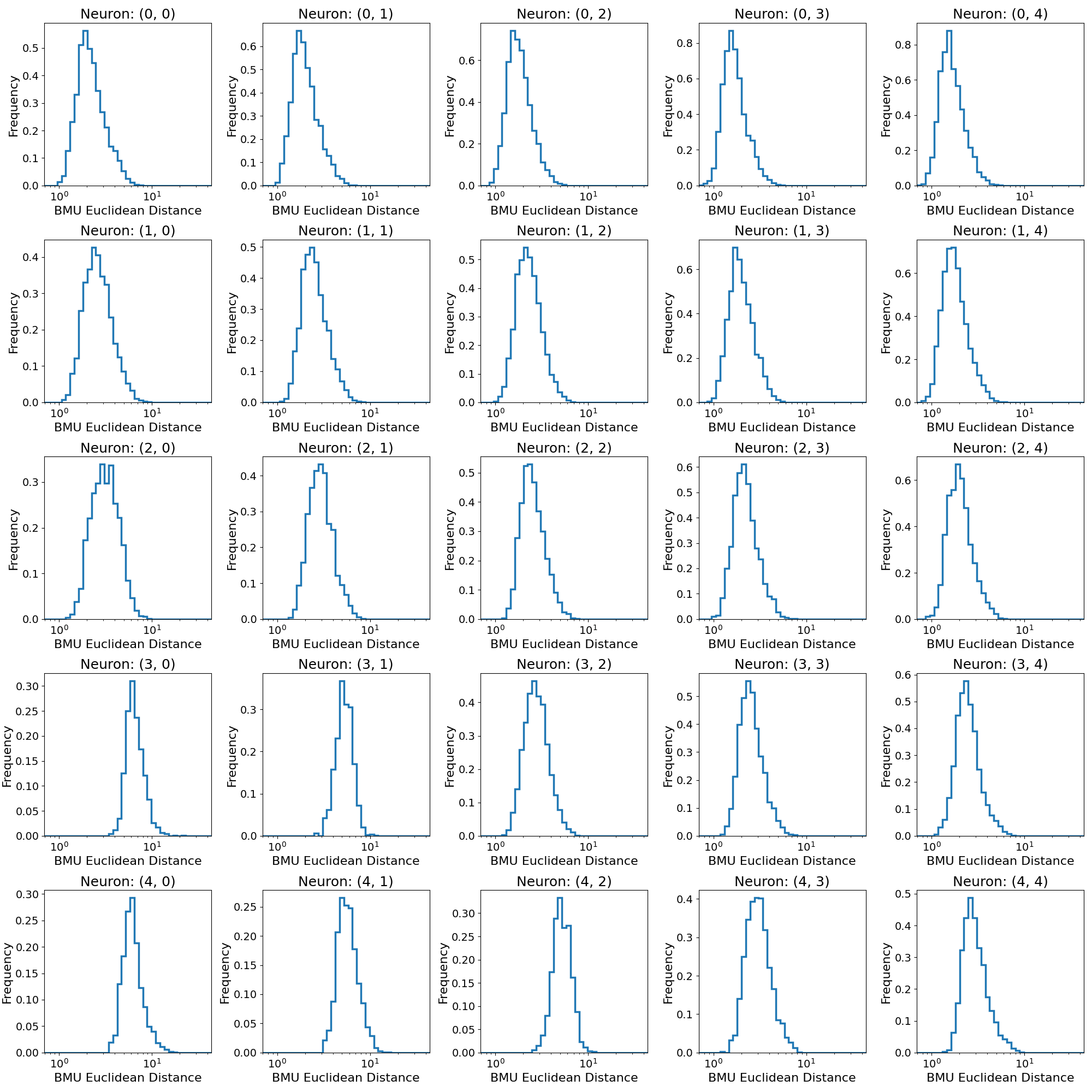}
        % \hspace{1mm}
    \caption{The distributions of the BMU Euclidean distance between an individual neuron in the SOM grid and all the input images for which it was chosen as the BMU for all the neurons in the top left quadrant of the SOM (Figure \ref{fig:som_final}).}
    \label{fig:top_left}
\end{figure*}

\begin{figure*}[ht]
    \centering
        \includegraphics[width=1.0\linewidth]{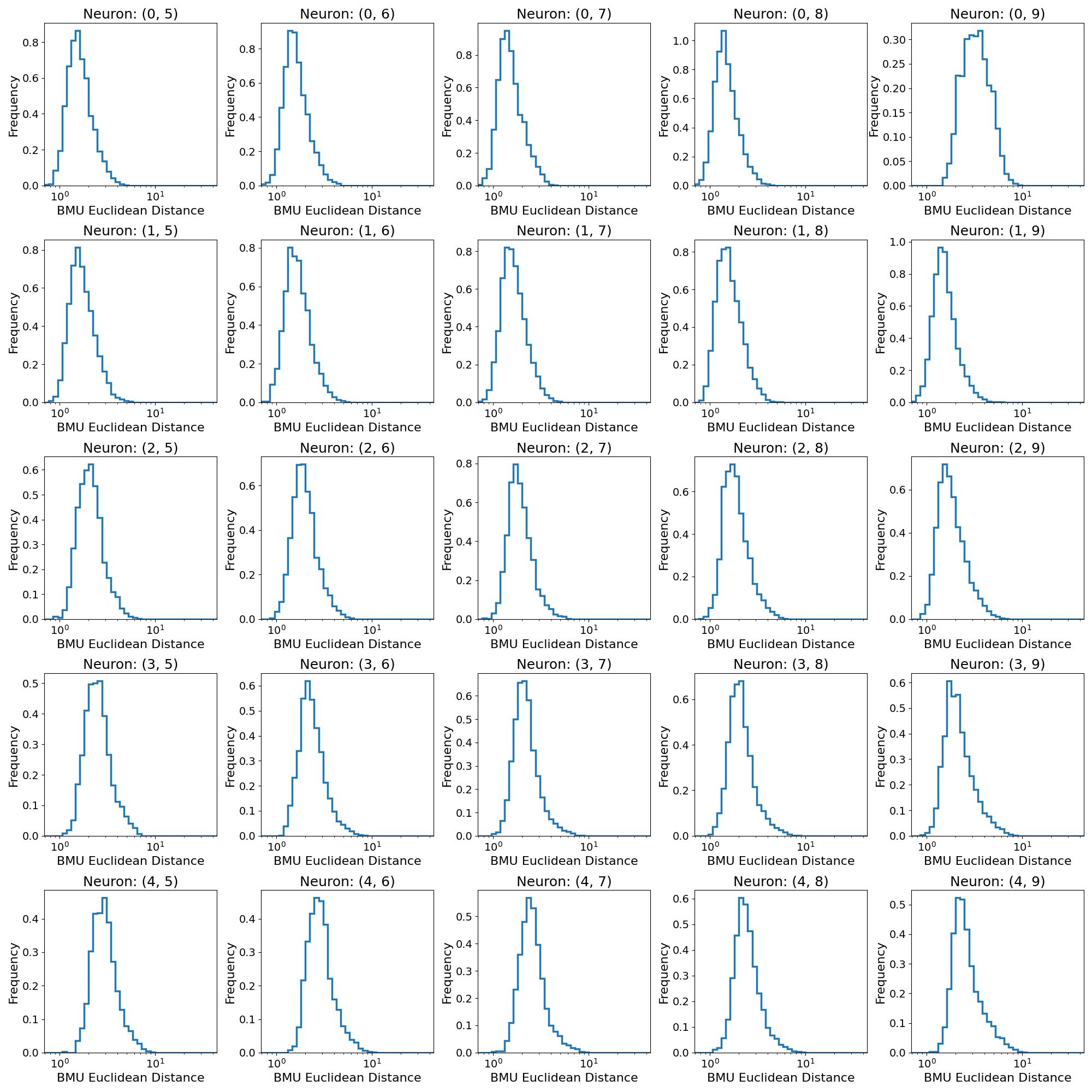}
        % \hspace{1mm}
    \caption{The distributions of the BMU Euclidean distance between an individual neuron in the SOM grid and all the input images for which it was chosen as the BMU for all the neurons in the top right quadrant of the SOM (Figure \ref{fig:som_final}).}
    \label{fig:top_right}
\end{figure*}

\begin{figure*}[ht]
    \centering
        \includegraphics[width=1.0\linewidth]{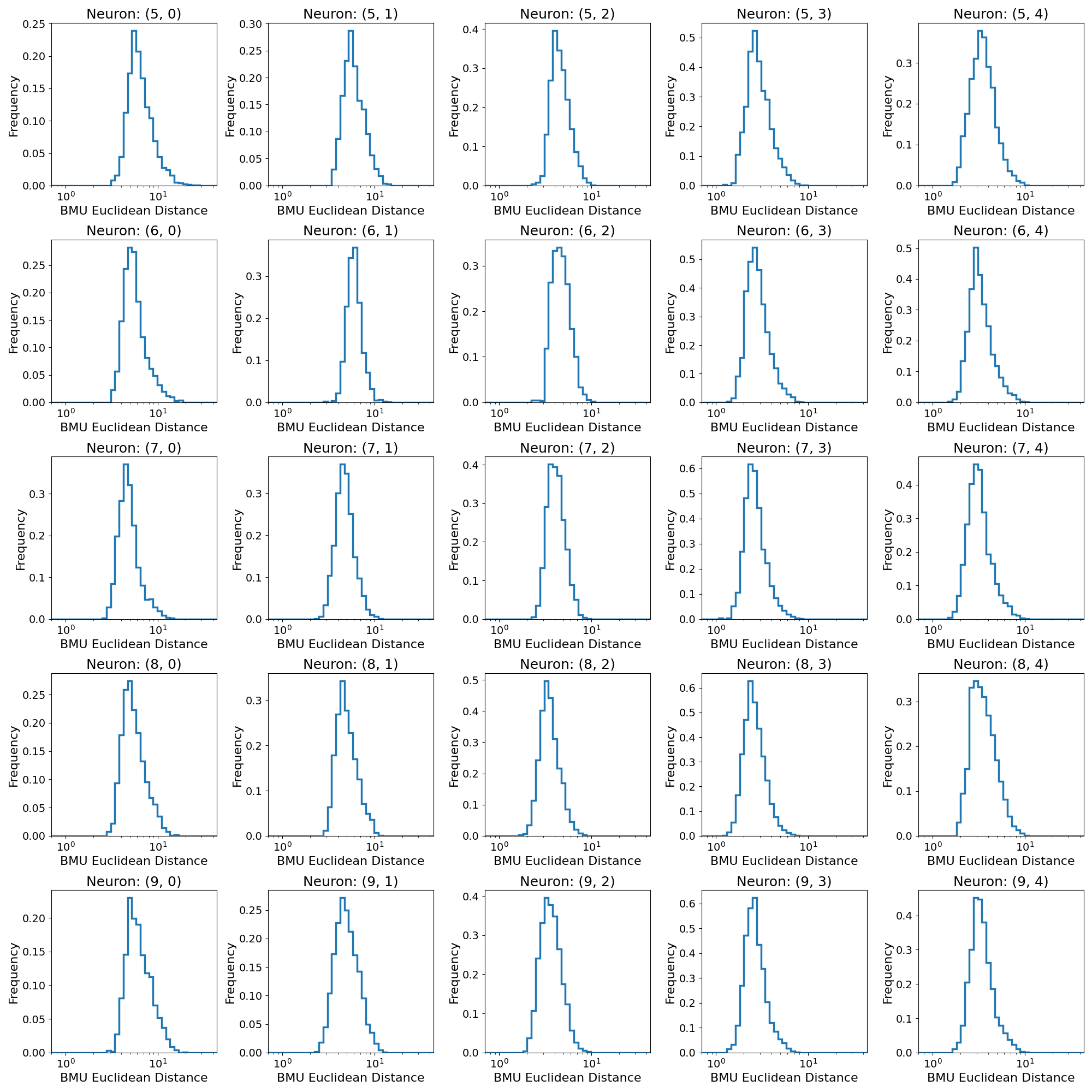}
        % \hspace{1mm}
    \caption{The distributions of the BMU Euclidean distance between an individual neuron in the SOM grid and all the input images for which it was chosen as the BMU for all the neurons in the bottom left quadrant of the SOM (Figure \ref{fig:som_final}).}
    \label{fig:bottom_left}
\end{figure*}

\begin{figure*}[ht]
    \centering
        \includegraphics[width=1.0\linewidth]{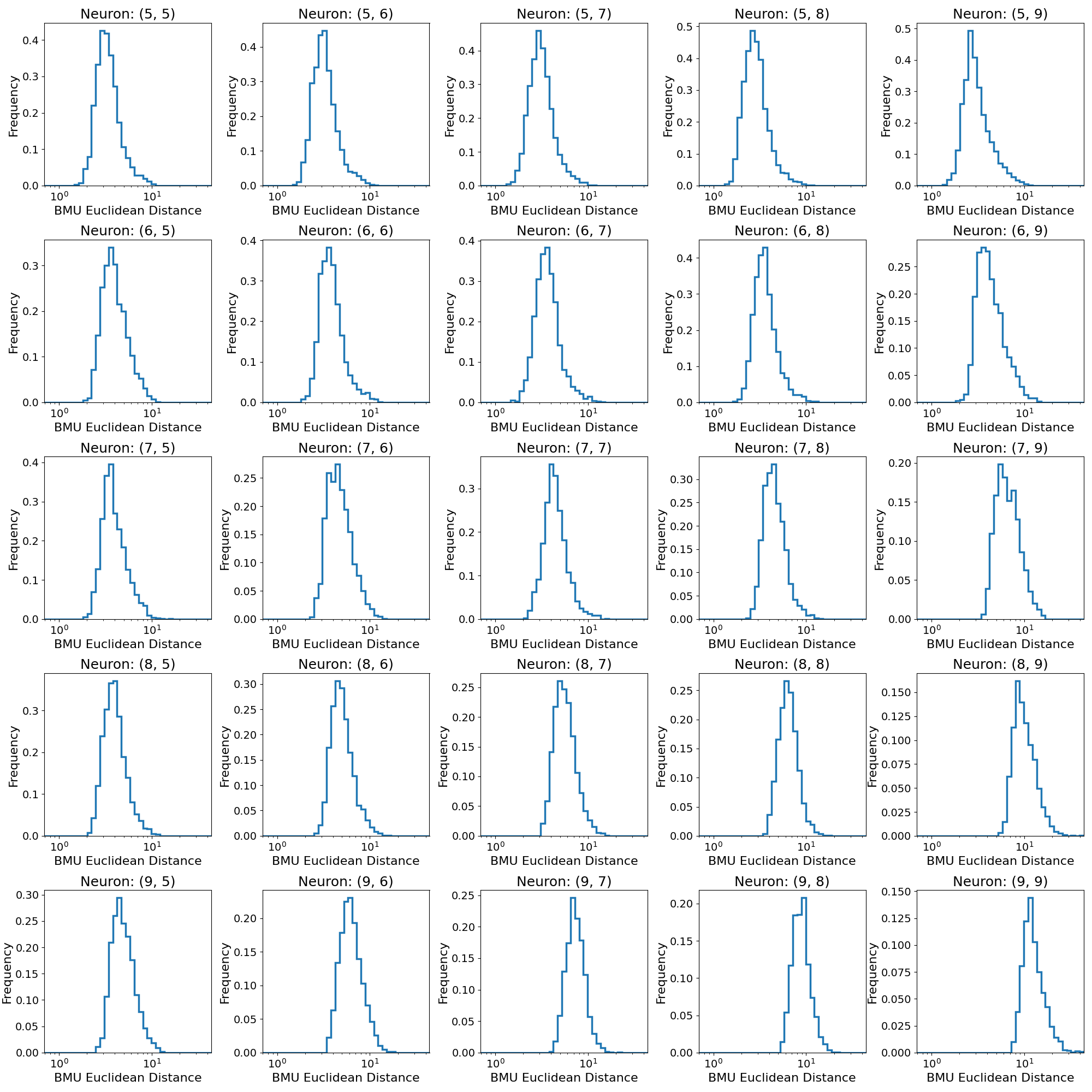}
        % \hspace{1mm}
    \caption{The distributions of the BMU Euclidean distance between an individual neuron in the SOM grid and all the input images for which it was chosen as the BMU for all the neurons in the bottom right quadrant of the SOM (Figure \ref{fig:som_final}).}
    \label{fig:bottom_right}
\end{figure*}

In Section \ref{inspection&mapping}, we have Figure \ref{fig:neuron_distance_log} which gives the distributions of the BMU Euclidean distance for the overall dataset. Here, we have included additional Figures \ref{fig:top_left}, \ref{fig:top_right}, \ref{fig:bottom_left}, and \ref{fig:bottom_right} which explore the distribution of the BMU Euclidean distance for individual neurons in the trained SOM grid divided into the four quadrants in the grid, i.e. the top left, top right, bottom left and bottom right (see the quadrants as indicated in red in Figure \ref{fig:som_final}). On the top quadrants, we mostly have smaller or simpler structures for which the Euclidean distances skews more towards the lower end. However, there is a trend for higher Euclidean distances in the bottom quadrants, especially the bottom right quadrant, and this can be attributed to there being larger or more extended sources in this SOM regions compared to the top left quadrant of the SOM. For these sources, we would expect a higher level of background and noise pixels, and their cutout size might also not be large enough to capture their full extent. As such, the higher Euclidean distances could be due to them not being modelled as well during SOM training. 

\end{document}